\documentclass[aps,prl,notitlepage,reprint,superscriptaddress,nofootinbib,longbibliography]{revtex4-1}

\usepackage{graphicx}
\usepackage{dcolumn}
\usepackage{bm}
\usepackage{amsmath}
\usepackage{upgreek}
\usepackage[colorlinks,linkcolor=red,citecolor=blue,urlcolor=red]{hyperref}
\usepackage[utf8]{inputenc}
\usepackage[T1]{fontenc}
\usepackage{mathptmx}
\usepackage[normalem]{ulem}
\usepackage[colorinlistoftodos]{todonotes}
\usepackage[mathlines]{lineno}
\renewcommand{\t}[1]{\mathrm{#1}}

\AtBeginDocument{%
	\newwrite\bibnotes
	\def\bibnotesext{Notes.bib}
	\immediate\openout\bibnotes=\jobname\bibnotesext
	\immediate\write\bibnotes{@CONTROL{REVTEX41Control}}
	\immediate\write\bibnotes{@CONTROL{%
			apsrev41Control,author="08",editor="1",pages="1",title="0",year="1"}}
	\if@filesw
	\immediate\write\@auxout{\string\citation{apsrev41Control}}%
	\fi
}%
\begin{document}
	
	\title{Sub-ppm Nanomechanical Absorption Spectroscopy of Silicon Nitride}
	
	\author{A. T. Land}
	\affiliation{Wyant College of Optical Sciences, University of Arizona, Tucson, AZ 85721, USA}
	
	\author{M. Dey Chowdhury}
	\affiliation{Wyant College of Optical Sciences, University of Arizona, Tucson, AZ 85721, USA}
	
	\author{A. R. Agrawal}%
	\affiliation{Wyant College of Optical Sciences, University of Arizona, Tucson, AZ 85721, USA}
	
	
	\author{D. J. Wilson}
	\email{dalziel@arizona.edu}
	\affiliation{Wyant College of Optical Sciences, University of Arizona, Tucson, AZ 85721, USA}
	
	\date{\today}
	\begin{abstract}
		Material absorption is a key limitation in nanophotonic systems; however, its characterization is often obscured by scattering and diffraction loss. Here we show that nanomechanical frequency spectroscopy can be used to characterize the absorption of a dielectric thin film at the parts-per-million (ppm) level, and use it to characterize the absorption of stoichiometric silicon nitride (Si$_3$N$_4$), a ubiquitous low-loss optomechanical material. Specifically, we track the frequency shift of a high-$Q$ Si$_3$N$_4$ trampoline resonator in response to photothermal heating by a $\sim10$ mW laser beam, and infer the absorption of the thin film from a model including thermal stress relaxation and both radiative and conductive heat transfer. A key insight is the presence of two thermalization timescales, a rapid ($\sim0.1$ sec) timescale due to radiative thermalization of the Si$_3$N$_4$ thin film, and a slow ($\sim100$ sec) timescale due to parasitic heating of the Si device chip. We infer the extinction coefficient of  Si$_3$N$_4$ to be $\sim0.1-1$ ppm in the 532 - 1550 nm wavelength range, comparable to bounds set by waveguide resonators and notably lower than estimates with membrane-in-the-middle cavity optomechanical systems. Our approach is applicable to a \mbox{broad variety of nanophotonic materials and may offer new insights into their potential.}
	\end{abstract}
	
	\maketitle
	
	Nanomechanical frequency shift spectroscopy has emerged as a powerful technique to characterize material absorption \cite{chien2018single, casci2019thin, kirchhof2023nanomechanical,west2023photothermal}.  Commonly termed ``nanomechanical absorption spectrocopy" (NAS), the technique involves monitoring the frequency shift of a nanomechanical resonator in response to laser heating of an absorber deposited on its surface.  Advantages of NAS include high sensitivity and compatibility with a variety of materials and optical frequencies.  Milestones and applications include single molecule imaging \cite{chien2018single}, femtowatt IR photodetection \cite{casci2019thin}, and 2D material spectroscopy~\cite{kirchhof2023nanomechanical}.
	
	In modeling NAS experiments, a common assumption is that absorption in the resonator material is negligible.  In practice, however, photothermal frequency shifts can be substantial even for low loss materials, provided that the nanomechanical resonator has sufficiently high quality factor ($Q$) or temperature sensitivity ($R$).  This is especially apparent in resonators made from high stress thin film materials such as silicon nitride, in which $Q$ is amplified by dissipation dilution \cite{sementilli2022nanomechanical} and $R$ is amplified by stress relaxation.  Such photothermal effects are indeed commonly utilized in the field of optomechanics for sensing applications and fundamental studies~\cite{usami2012optical,zhu2016plasmonic}.
	
	Since material absorption characterization is a crucial consideration for nanophotonics, it is intriguing to consider NAS as an alternative to conventional techniques such as cavity absorption spectrocopy (CAS) \cite{armani2006heavy,nitkowski2008cavity}, in which low level absorption can be obscured by scattering and diffraction loss.  Obstacles to this approach are at least two-fold:  first, absorption must be inferred from a photothermal frequency shift model that includes both extensive and intensive properties of the nanomechanical resonator; second---as we encounter below---the frequency shift must be sufficiently time resolved, to distinguish (rapid) local heating of the nanomechanical resonator from (slow) parasitic heating of its surroundings.
	
	Armed with these expectations, here we describe an experiment in which NAS is used to estimate the extinction coefficient $\kappa$ of stoichiometric silicon nitride (Si$_3$N$_4$), a ubiquitous low loss material in nanophotonics and optomechanics.  Using a high stress Si$_3$N$_4$ trampoline resonator with an enhanced $Q R$ product, thermal-noise-limited interferometric readout, and a semi-analytical model for radiative and conductive heat transfer, we infer $\kappa\sim 0.1 - 1$ ppm in the 532 - 1550 nm wavelength range, comparable to previous bounds inferred from waveguide resonators, but notably lower than previous estimates made with membrane-in-the-middle cavity optomechanical systems.  While subject to refinement, these results suggest that NAS holds promise as an alternative to CAS for absorption spectroscopy of \mbox{low loss nanophotonics materials. }
	
	
	\begin{figure}[b!]
		\vspace{-2mm}
		\centering  \includegraphics[width=1\columnwidth]{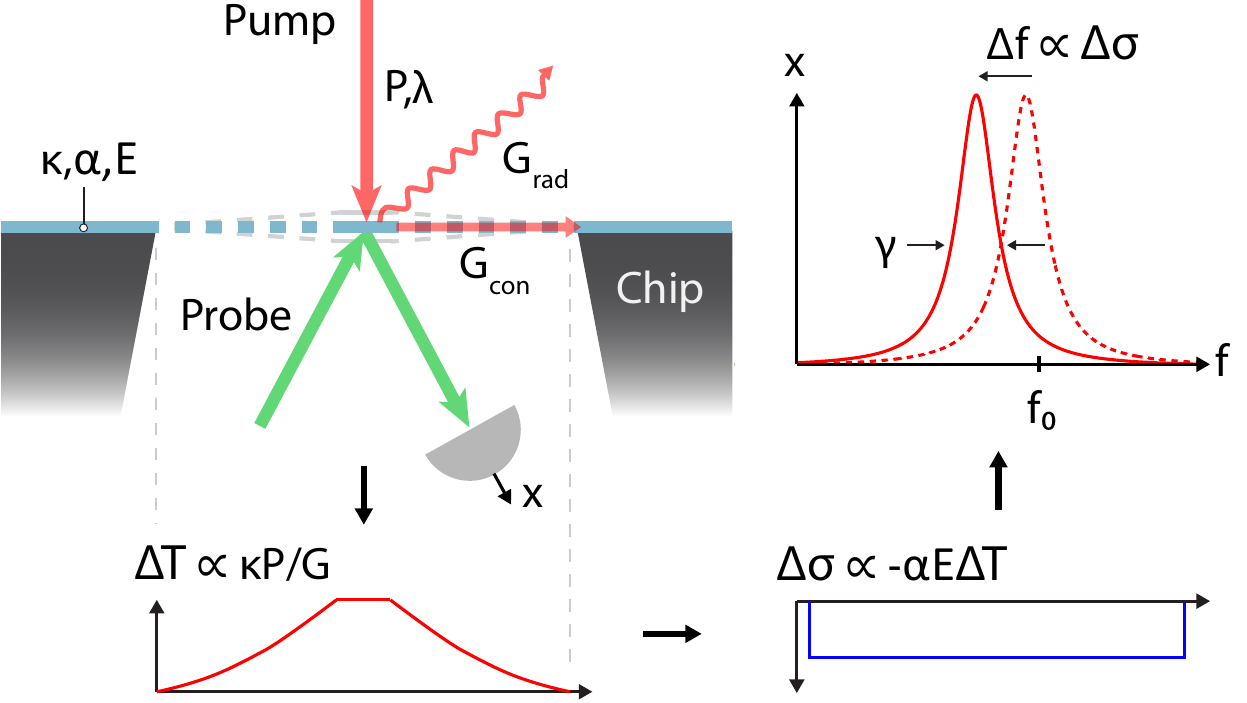}
		\caption{Overview of nanomechanical absorption spectroscopy.  Light from a pump laser (power $P$, wavelength $\lambda$) irradiates a nanomechanical resonator.  Absorption produces a temperature shift $\Delta T$, which in turns gives rise to a stress change $\Delta \sigma$ and a mechanical frequency shift $\Delta f$, as a function of material and geometric properties (see main text).   The frequency shift is monitored using an auxiliary probe laser.}
		\label{fig:1}
	\end{figure}

	\begin{figure*}[ht!]
		\centering  \includegraphics[width=2\columnwidth]{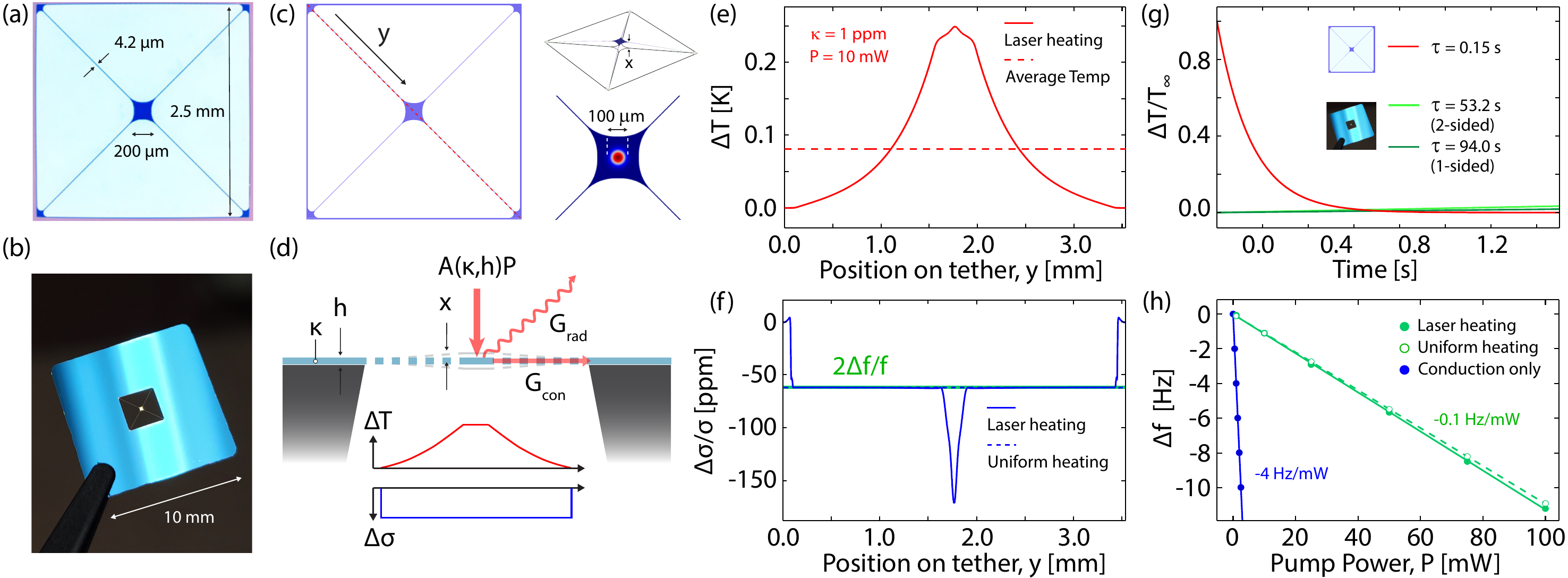}
		\caption{Device and numerical simulations. (a) Inverted-color microscope image of the trampoline device.  (b) Photograph of the device chip.  (c) Left: Geometry and cross-section of the numerical simulation. Top right: Simulated mechanical mode shape. Bottom right: Pad overlaid with simulated Gaussian laser beam profile. (d) Sketch of the photothermal heat transfer problem: Absorbed pump power $A P$ (film absorbance $A$) gives rise to a temperature change $\Delta T$ when balanced against radiative $G_\t{rad}$ and conductive $G_\t{con}$ heat transfer.  (e) Simulated temperature rise profile and average temperature rise $\overline{\Delta T}$ for $P = 10\,\t{mW}$ and $\kappa = 1$ ppm. (f) Simulated fractional stress change due to localized heating and equivalent uniform heating by $\overline{\Delta T}$. Overlaid is the fractional frequency shift using Eq. \ref{eq:3} and $\eta = 2.8$. (g) Simulated thermalization time of the trampoline pad and device chip.  (h) Simulated steady-state frequency shift versus $P$ for $\kappa = 1$ ppm.}\label{fig:2}
		\vspace{-2mm}
	\end{figure*}
	
	An overview of our approach is shown in Fig. \ref{fig:1}.  As illustrated at top left, the trampoline resonator is illuminated on its central pad by two laser beams: a weak ``probe'' beam used to monitor the resonator frequency and a strong ``pump" beam used to induce photothermal heating $\Delta T$.  In a lumped mass model, the resulting frequency shift can be expressed as \cite{west2023photothermal} 
	\begin{equation}\label{eq:1}
		\frac{\Delta f}{f_0}  \approx \frac{1}{f_0}\frac{\partial f}{\partial \sigma}\cdot\frac{\partial \sigma}{\partial T}\cdot \overline{\Delta T}\approx\left(\frac{1}{2\sigma}\right)\left( -\alpha E\right)\left(\frac{4\pi h}{\lambda}\frac{P\kappa}{G}\frac{\beta}{\eta}\right),
	\end{equation}
	where $f_0 \propto \sqrt{\sigma}$ is the ``cold" ($P = 0$) resonance frequency; $\overline{\Delta T}$, $\sigma$, $\alpha$, $E$, and $h$ are the average temperature rise, tensile stress, thermal expansion coefficient, and Young's modulus of the released film, respectively; $P$ and $\lambda$ are the pump power and wavelength, respectively; $G$ is the effective thermal conductance of the pad; $\beta\approx 1$ is a unitless constant accounting for thin film interference; and $\eta\sim 1$ is a unitless constant depending on the temperature distribution. As a figure of merit, we define the NAS spectral resolution as the extinction coefficient necessary to shift the resonance by one linewidth ($f_0/Q\equiv \gamma $)
	\begin{equation}\label{eq:2}
		\kappa_\gamma \approx \frac{2\sigma}{\alpha E}\frac{\lambda}{4\pi h}\frac{G}{Q P}\frac{\eta}{\beta}.
	\end{equation}
	As discussed below, we find that $\kappa_\gamma$ can reach the ppm level ($10^{-6}$) for milliwatts of pump power $P$ and $\lambda\sim 1\,\mu\t{m}$.

	In our experiment, we used a Si$_3$N$_4$ trampoline resonator with typical dimensions \cite{reinhardt2016ultralow,pluchar2020towards}  shown in Fig. 2a. We note that Si$_3$N$_4$ thin film resonators (strings and membranes) have been widely explored in cavity optomechanics \cite{aspelmeyer2014cavity} because their high-stress ($\sigma\approx 1$ GPa, produced by thermal mismatch with the Si chip during chemical vapor deposition) results in high $Q$ due to dissipation dilution \cite{sementilli2022nanomechanical}.  By contrast, Si-rich SiN$_x$ thin film resonators are commonly used in NAS, because their higher absorption and lower stress ($\sigma\approx 0.1$ MPa) results in a larger fractional photothermal frequency shift  according to Eq.~\ref{eq:1}.  In either case, the trampoline geometry is motivated by the high tether aspect ratio, which enables higher $Q/(m f_0^2)$ (effective mass $m$) and $Q/G$ ratios---and therefore higher force and photothermal sensitivity---respectively. 
	
	Inferring the extinction coefficient $\kappa$ from the photothermal frequency shift requires modeling the temperature rise $\overline{\Delta T}(\kappa,P)$ and frequency response $\partial f/\partial T$.  The lumped mass model in Eq.~\ref{eq:1} assumes the average temperature rise of the trampoline obeys Ohm's law $\overline{\Delta T} = PA/G$, where $A\approx 4\pi h\kappa\beta/\lambda$ is the film absorbance.  The thermal conductance from the pad $G$ is approximated by the sum of the conductance of the four tethers (length $L$, width $w$) $G_\t{con} = 4kwh/L$ and the radiative conductance of the pad (width $w_\t{p}$) $G_\t{rad} \approx 8w_\t{p}^2\sigma_\t{B}\epsilon T_0^3$ \cite{zhang2020radiative}, where $\sigma_\t{B}$, $\epsilon$, and $T_0$ are Boltzmann's constant, the film emissivity, and the initial film temperature, respectively.  The resulting frequency shift can be heuristically derived from the functional dependence $f_0\propto \sqrt{\sigma}$ and the stress-strain relation $\Delta \sigma = E\Delta L/L = -\alpha E\overline{\Delta T}$, where the minus sign comes from the assumption that the trampoline dimensions are constrained by the chip \cite{zhang2020radiative}. Notably, for our device dimensions---$L \approx 1.6\,\t{mm}$, $w\approx 4.2\,\mu\t{m}$, $w_p\approx 200\,\mu\t{m}$, and $h\approx 75\,\mu\t{m}$---and the material properties in Table 1, radiation dominates conductive heat transfer by an order of magnitude ($G_\t{rad}\approx 20 G_\t{con}$) \cite{zhang2020radiative,piller2020thermal}, motivating the following approximate model for the fractional frequency shift:
	\begin{equation}\label{eq:3}
		\frac{\Delta f}{f_0} \approx - \frac{\pi\alpha E h P \kappa}{4\sigma w_\t{p}^2\sigma_\t{B}\epsilon T_0^3 \lambda}\frac{\beta}{\eta}\left(1+\frac{G_\t{con}}{G_\t{rad}}\right)^{-1}.
	\end{equation}
	As shown in Fig. \ref{fig:2}, we confirmed this model by carrying out a finite element (COMSOL) simulation. 
	We find that the simulated $\Delta f/f_0$ agrees with Eq. \ref{eq:3} using $\eta \approx 2.8$, which differs by $40\%$ from the value $\eta = 2$ for purely conductive case $G_\t{con}\gg G_\t{rad}$ \cite{west2023photothermal}.  This discrepancy is due to the nonlinear temperature profile in the radiation dominated case (Fig. \ref{fig:2}e), which leads to a reduction of $ \overline{\Delta T}/\Delta T_\t{max} \approx \eta^{-1}$.
	\begin{table}[b!]
		\begin{ruledtabular}
			\begin{tabular}{ l l c c }
				Quantity & Value & Units & Reference\\
				\colrule
				Tensile Stress ($\sigma$) & 0.85 & GPa & \cite{pluchar2020towards}\\
				Young's Modulus (E) & 250[50] & GPa & \cite{klass2022determining,zhang2006bulk}\\
				Thermal Expansion Coeff. ($\alpha$) & $2.0[0.4] \times 10^{-6}$ & $\t K^{-1}$& \cite{zhang2020radiative,snell2022heat}\\
				Thermal Conductivity (k) & 3.0[0.2] & $\frac{\t W}{\t m \cdot K}$& \cite{zhang2020radiative,piller2020thermal}\\
				Emissivity ($\epsilon$) & $0.09$ & - & \cite{zhang2020radiative,zhang2021erratum,piller2020thermal}\\
				Density ($\rho$) & 2900[200] & $\frac{\t kg}{\t m^3}$& \cite{piller2020thermal,vivekananthan2020primary}\\
				Specific Heat (c) & 750[50] & $\frac{\t J}{\t kg \cdot K}$& \cite{ftouni2015thermal,kuwabara2008lattice}
			\end{tabular}
			\caption{Physical constants used in this work.  Values in brackets represent variation over cited references. Except for emissivity, all values are for thin-film stoichiometic (Si$_3$N$_4$) silicon nitride.}
		\end{ruledtabular}
		
		\label{table:1}
	\end{table}
	
	\begin{figure*}[ht!]
		\vspace{-1mm}
		\centering  \includegraphics[width=2\columnwidth]{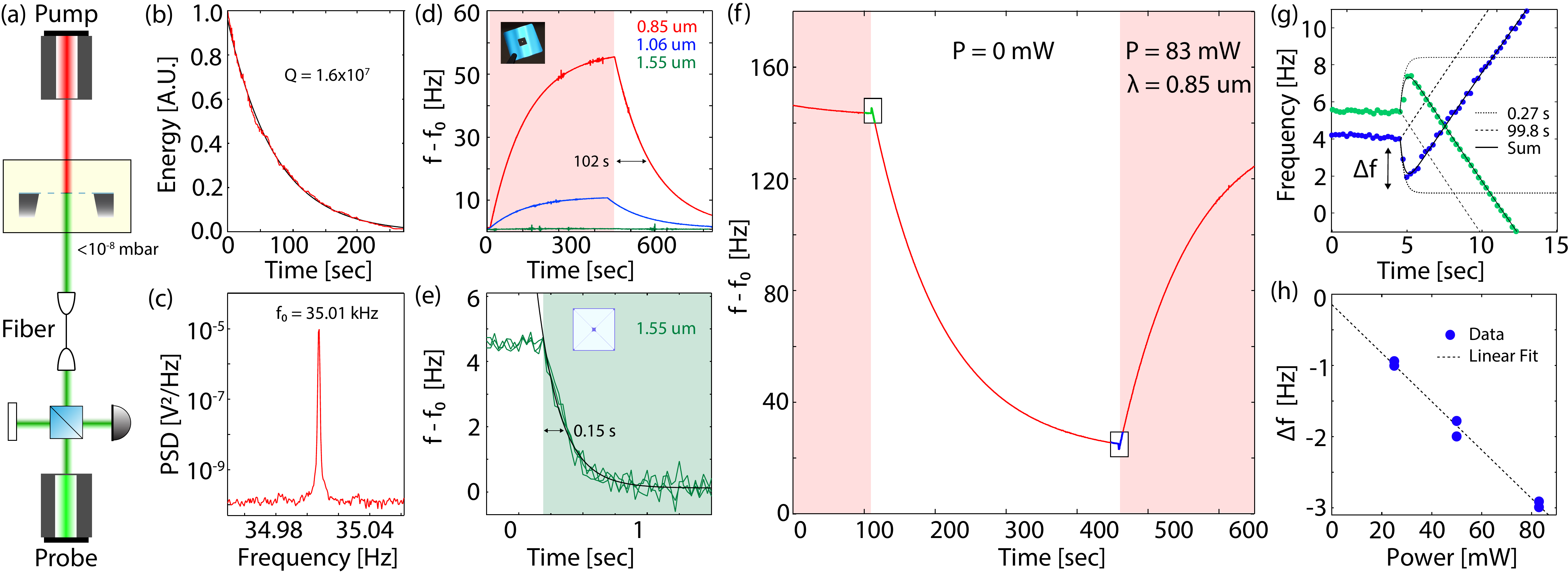}
		\caption{Experimental setup and absorption measurement. (a) Sketch of the experimental apparatus.  The trampoline is housed in an ultra-high-vacuum chamber ($<10^{-8}$ mbar).  Counter-propagating pump and probe lasers irradiate the trampoline pad through opposite viewports. 
			The reflected probe field is integrated into a Michelson interferometer \cite{pluchar2020towards}, enabling the displacement of the trampoline to be monitored.  (b) Ringdown of the $f_0 \approx 35$ kHz fundamental trampoline mode for the device in Fig. 2 (for details, see \cite{pluchar2020towards}). (c) Thermal-noise-limited displacement measurement with a probe power of 1 mW, in power spectral density (PSD) units, referred to the voltage input of the digitizer (National Instruments PXI 4461). (d)  Response of the trampoline frequency $f$ to toggling on (opaque) and off (white) a 2 mW beam aligned on the Si device chip, for three wavelengths.  (e) Response of $f$ to toggling a 30 mW, 1550 nm beam aligned on the trampoline pad, using a digitizer RBW of 40 Hz. (f)  Response of $f$ to toggling a 83 mW, 850 nm beam aligned on the trampoline pad, using an RBW of 10 Hz.  (g) 
			Magnification of the boxed regions in (f), showing the rapid initial frequency shift due to trampoline heating and the fitting routine used to distill the steady state shift $\Delta f$ from slow drift due to chip heating.  (h) Plot of $\Delta f$ versus $P$ for $\lambda = 850$ nm, confirming linearity. }\label{fig:3}
		\vspace{-1mm}
	\end{figure*}
	
	In addition to modeling the photothermal response, we have found that it is important to consider the thermalization timescale of both the nanomechanical resonator and the device chip---the latter of which is parasitically heated (possibly due to scattered light) when the pad is illuminated, resulting in a frequency shift of opposite sign.  In a lumped mass model of the trampoline (device chip), the timescales are approximated by $\tau = C/G$, where $C = C_V V$, $C_V$ and $C$ are the heat capacity, volumetric heat capacitance, and thermal conductance out of the trampoline pad (device chip), respectively. In Fig. \ref{fig:2}g, we present finite element simulations of the thermalization times for our device using $\epsilon = 0.09$ \cite{zhang2020radiative,zhang2021erratum} and $0.6$ for the Si$_3$N$_4$ trampoline and Si chip, respectively.  The predicted value for the trampoline, $\tau= 0.15$ sec, agrees well with measurements described below (see Fig \ref{fig:3}e), and corroborates our use of low-stress (Si rich) silicon nitride emissivity values \cite{zhang2020radiative,zhang2021erratum} in analysis of high-stress Si$_3$N$_4$ thin films.  The prediction for a one-sided device chip,  $\tau = 94$ sec, likewise agrees well with measurements (see Fig. \ref{fig:3}d), consistent with the opaque underside of the sample holder.  
	
	Details of the experiment and trampoline characterization are shown in Fig. \ref{fig:3}a-c.  To minimize gas damping, the device is housed in an ultra-high vacuum ($<10^{-8}\,\t{mbar}$) chamber with optical access in two directions.  From one direction, the trampoline is probed in reflection with a relatively weak ($\sim 100$ $\t\mu$W), 850 nm probe beam derived from a low noise external cavity diode laser (Newport Velocity TLB-6716), enabling low noise displacement readout via Michelson interferometry (for details see \cite{pluchar2020towards}).  From the opposite direction, the trampoline pad is irradiated with a strong ($1 - 100$ mW), focused ($< 100\,\mu\t{m}$ diameter) pump beam derived from an assortment of lasers operating at different wavelengths: 532 nm (Laser Quantum Finesse), 633 nm (Coherent He-Ne), 780 - 950 nm (M-Squared Solstis Ti-Sapphire), 1064 nm (AMOCO YAG), and 1550 nm (EMCORE 1752A passed through an Erbium-doped fiber amplifier).  Ringdown and thermal-noise-limited displacement measurements of the fundamental trampoline mode are shown in Fig. \ref{fig:3}b,c, revealing $Q=1.6\times 10^7$ and $f_0\approx 35$ kHz. As highlighted in Fig. \ref{fig:3}c, the high $Q/m$ of the trampoline enables thermal noise to be resolved with a FFT resolution bandwidth greater than 10 Hz, enabling frequency tracking within the $\sim 0.1\,\t{sec}$ thermalization time.
	
	\begin{figure*}[ht!]
		\vspace{-1.8mm}
		\centering  \includegraphics[width=2\columnwidth]{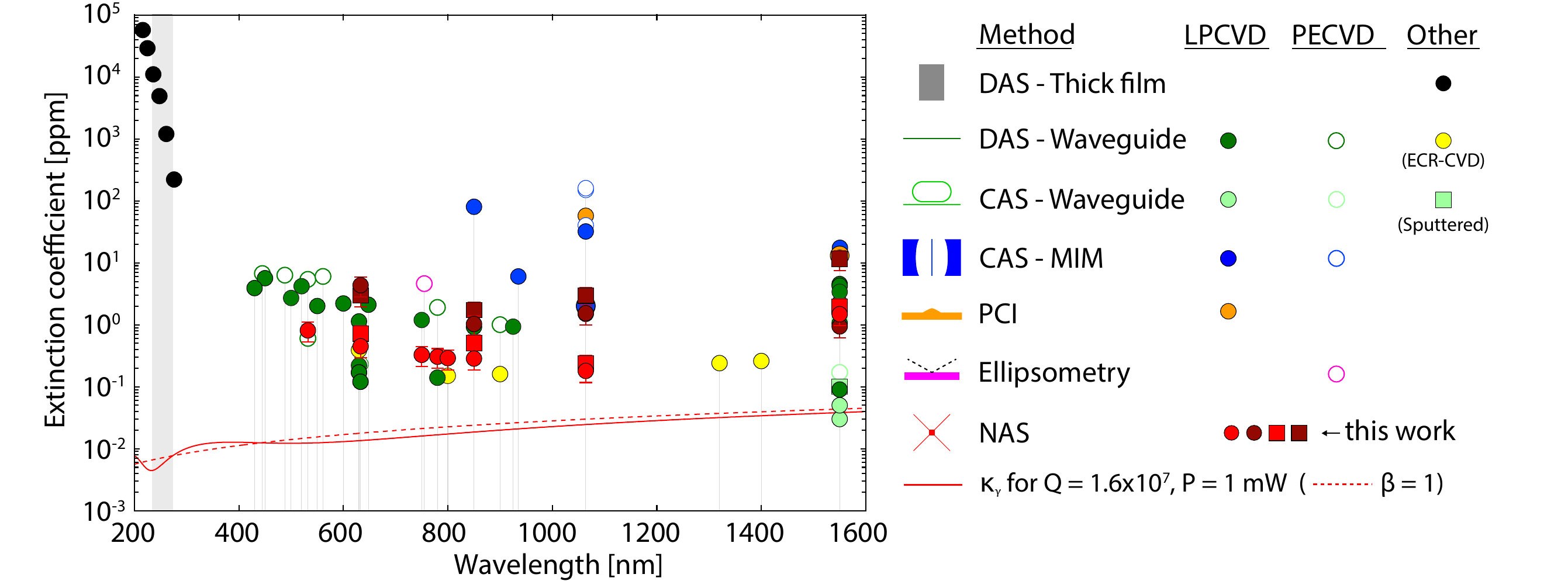}
		\caption{Compilation of extinction coefficient measurements for silicon nitride. 
			Surveyed methods include direct (single-pass) absorption spectroscopy (DAS) in thick films \cite{philipp1973optical} and high-confinement ($>50\%$) waveguides \cite{bauters2011ultra,xiong2020sin,wang2018nonlinear,daldosso2004comparison,bulla1999design,inukai1994optical,daldosso2004fabrication,buzaverov2023low,smith2023sin,yong2022power,sacher2019visible,subramanian2013low}, cavity-enhanced absorption spectroscopy (CAS) with waveguide microresonators \cite{liu2021high,ji2023ultra,zhang2023room} and membrane-in-the-middle (MIM) cavity optomechanical systems \cite{karuza2012tunable,wilson2009cavity,wilson2012cavity,serra2016microfabrication,sankey2010strong,weaver2017coherent,stambaugh2014cavity}, photothermal common path interferometry (PCI) \cite{steinlechner2017optical}, ellipsometry \cite{nejadriahi2020thermo}, and nanomechanical frequency shift absorption spectroscopy (NAS, this work).  The gray region is the rough cut-off wavelength for two-photon absorption.  Red curves are the spectral resolution (Eq. \ref{eq:2}) of our device with $P = 1$ mW.}\label{fig:4}
		\vspace{-1mm}
	\end{figure*}
	
	We now describe a typical photothermal frequency shift measurement, shown in Fig. \ref{fig:3}f.  For this measurement, a $\lambda= 850$ nm, $P=83$ mW pump beam was shuttered on and off with a beam block; meanwhile, the frequency of the  fundamental trampoline mode was tracked using a Labview-based fitting routine applied to displacement measurements digitized at 10 Hz.  As is evident from the data, two frequency shifts are observed when the pump is shuttered on: a small, rapid, negative shift and a large, slow, positive shift, with exponential settling times of $\tau\approx 0.2$ s and $\tau\approx100$ s, respectively.  The sign of these shifts and their qualitative agreement with predicted thermalization times (Fig. 2g) identify them with photothermal heating of the trampoline and device chip, respectively.  Fitting to a double exponential, as shown in Fig. 3g, implies a frequency shift of $\Delta f = 3$ Hz due to heating of the trampoline.  Averaging $\sim 30$ of such measurements and performing a power sweep to  confirm linearity (Fig. \ref{fig:3}h), we infer a frequency-power sensitivity of $\Delta f/P = 32.5$ Hz/W, corresponding to an extinction coefficient of $\kappa = 0.28\pm0.10$ ppm. The $34\%$ standard deviation is obtained from material properties in Table 1 ($31\%$), variance across the measurement ensemble ($10\%$), and an estimated power uncertainty ($10\%$).  
	
	
	Figure \ref{fig:4} shows the main result of this Letter: an estimate of the extinction coefficient of thin film Si$_3$N$_4$ from 532 nm to 1550 nm, using Eq. \ref{eq:1}, measurements as shown in Fig. \ref{fig:3}, and the series of pump lasers listed above. A total of four devices were measured: two identical to the device in Fig. \ref{fig:2}, and two from a different wafer and LPCVD run, with narrower (2.1 $\mu$m), longer (2.0 mm) tethers and a simulated $\eta = 4.2$.  Special care was taken to identify samples with as few particles (e.g. dust or photoresist) on the pad and tethers as possible, noting that these ``hotspots" produced large, alignment sensitive absorption, consistent with dedicated work on NAS particle sensing.  As indicated with solid red circles and squares at $\lambda$ = (532, 633, 750, 780, 800, 850, 1064, 1550) nm, we obtain minimum values of $\kappa$ = (0.81, 0.45, 0.33, 0.30, 0.29, 0.28, 0.18, 0.94) using the Fig. 1 device (except for the the value at 1550 nm, which was obtained with the second device type). All other devices exhibited systematically higher apparent absorption, in $\kappa\sim 1$ ppm range.  A common trend of decreasing absorption with wavelength was observed, except at 1550 nm, where a systematic rise was observed.  A comparison is made with results for thick and thin film silicon nitride using a variety of techniques, for both stoichiometric (Si$_3$N$_4$) and Si rich (SiN$_x$) deposited by LPCVD, PECVD and other methods.
	
	As evident in Fig. \ref{fig:4}, the thin-film Si$_3$N$_4$ extinction coefficients we infer from NAS, $\kappa \sim 0.1 - 1$ ppm (corresponding to a propagation loss of $\sim 0.1 - 1$ dB/cm), are on par with lower bounds from previous works---specifically, those obtained using the scattering detection method on $\sim 6\,\mu\t{m}$ thick films \cite{inukai1994optical} and by transmission measurements (DAS) on 200-nm-thick channel waveguides \cite{daldosso2004fabrication}.  They are slightly lower than common results obtained using waveguide-based DAS/CAS \cite{bauters2011ultra,xiong2020sin,liu2021high,wang2018nonlinear,ji2023ultra,daldosso2004comparison,bulla1999design,inukai1994optical,daldosso2004fabrication,zhang2023room,buzaverov2023low,smith2023sin,yong2022power,sacher2019visible,subramanian2013low}, and significantly lower than obtained by techniques applied to Si$_3$N$_4$ membranes, such as photothermal common path interferometry (PCI) \cite{steinlechner2017optical} and membrane-in-the-middle (MIM) CAS \cite{karuza2012tunable,wilson2009cavity,wilson2012cavity,serra2016microfabrication,sankey2010strong,weaver2017coherent,stambaugh2014cavity}.  A possible reason for this discrepancy is the susceptibility of thin film DAS and CAS to scattering and diffraction loss, including sidewall roughness for waveguides and tip-tilt misalignment in MIM systems.  We also re-emphasize that NAS is sensitive to uncertainties in material properties entering Eq. \ref{eq:1}. However, these uncertainties don't appear to account for the 10-fold discrepancy with MIM and PCI measurements. On the other hand, as highlighted by the red curves in Fig. \ref{fig:4}, Eq. \ref{eq:2} implies that the linewidth-equivalent resolution $\kappa_\gamma$ for our Si$_3$N$_4$ trampoline is well below 0.1 ppm level, suggesting that the inferred absorption levels is at least a priori accessible with high resolution.
	
	Looking forward, it will be important to probe the sensitivity, accuracy, and material independence of NAS as a thin film absorbance probe. Towards this end, we first note that frequency metrology of nanomechanical resonators is a mature discipline, with fractional Allan deviations $\sigma_y<Q^{-1}$ routinely achieved, including $\sigma_y<10^{-7}$ for Si$_3$N$_4$ nanostrings and membranes \cite{sadeghi2020frequency,zhang2023demonstration}; these experiments can be readily repurposed for substrate absorption estimates.  With regards to accuracy, our next step will be to reapply our method to low-stress, PECVD SiN$_x$ membranes for which the absorption is known to significantly larger, as well as with alternative membrane or string geometries, for which conductive heat transfer can dominate over radiative heat transfer, eliminating variables such as emissivity of the thin film.  Finally, NAS seems well-suited as an absorbance probe for a wide variety of optomechanics materials, particularly those grown naturally under tensile stress, such as strained-Si \cite{beccari2022strained}, GaAs \cite{liu2011high}, InGaP \cite{cole2014tensile}, and SiC \cite{xu2023high}.

	\section*{Acknowledgments}
	
	The authors thank Christian Pluchar and Charles Condos for assisting in developing the experimental apparatus. This work was supported by NSF Grant ECCS-1945832. ATL acknowledges scholarship support from the Galileo Circle Organization. ARA acknowledges support from an iGlobes CNRS-UArizona fellowship. Finally, the reactive ion etcher used for this study was funded by an NSF MRI grant, ECCS-1725571.
	
	\bibliography{ref}

\begin{thebibliography}{52}%
\makeatletter
\providecommand \@ifxundefined [1]{%
 \@ifx{#1\undefined}
}%
\providecommand \@ifnum [1]{%
 \ifnum #1\expandafter \@firstoftwo
 \else \expandafter \@secondoftwo
 \fi
}%
\providecommand \@ifx [1]{%
 \ifx #1\expandafter \@firstoftwo
 \else \expandafter \@secondoftwo
 \fi
}%
\providecommand \natexlab [1]{#1}%
\providecommand \enquote  [1]{``#1''}%
\providecommand \bibnamefont  [1]{#1}%
\providecommand \bibfnamefont [1]{#1}%
\providecommand \citenamefont [1]{#1}%
\providecommand \href@noop [0]{\@secondoftwo}%
\providecommand \href [0]{\begingroup \@sanitize@url \@href}%
\providecommand \@href[1]{\@@startlink{#1}\@@href}%
\providecommand \@@href[1]{\endgroup#1\@@endlink}%
\providecommand \@sanitize@url [0]{\catcode `\\12\catcode `\$12\catcode `\&12\catcode `\#12\catcode `\^12\catcode `\_12\catcode `\%12\relax}%
\providecommand \@@startlink[1]{}%
\providecommand \@@endlink[0]{}%
\providecommand \url  [0]{\begingroup\@sanitize@url \@url }%
\providecommand \@url [1]{\endgroup\@href {#1}{\urlprefix }}%
\providecommand \urlprefix  [0]{URL }%
\providecommand \Eprint [0]{\href }%
\providecommand \doibase [0]{http://dx.doi.org/}%
\providecommand \selectlanguage [0]{\@gobble}%
\providecommand \bibinfo  [0]{\@secondoftwo}%
\providecommand \bibfield  [0]{\@secondoftwo}%
\providecommand \translation [1]{[#1]}%
\providecommand \BibitemOpen [0]{}%
\providecommand \bibitemStop [0]{}%
\providecommand \bibitemNoStop [0]{.\EOS\space}%
\providecommand \EOS [0]{\spacefactor3000\relax}%
\providecommand \BibitemShut  [1]{\csname bibitem#1\endcsname}%
\let\auto@bib@innerbib\@empty
\bibitem [{\citenamefont {Chien}\ \emph {et~al.}(2018)\citenamefont {Chien}, \citenamefont {Brameshuber}, \citenamefont {Rossboth}, \citenamefont {Sch{\"u}tz},\ and\ \citenamefont {Schmid}}]{chien2018single}%
  \BibitemOpen
  \bibfield  {author} {\bibinfo {author} {\bibfnamefont {M.-H.}\ \bibnamefont {Chien}}, \bibinfo {author} {\bibfnamefont {M.}~\bibnamefont {Brameshuber}}, \bibinfo {author} {\bibfnamefont {B.~K.}\ \bibnamefont {Rossboth}}, \bibinfo {author} {\bibfnamefont {G.~J.}\ \bibnamefont {Sch{\"u}tz}}, \ and\ \bibinfo {author} {\bibfnamefont {S.}~\bibnamefont {Schmid}},\ }\bibfield  {title} {\enquote {\bibinfo {title} {Single-molecule optical absorption imaging by nanomechanical photothermal sensing},}\ }\href {https://www.pnas.org/doi/abs/10.1073/pnas.1804174115} {\bibfield  {journal} {\bibinfo  {journal} {Proceedings of the National Academy of Sciences}\ }\textbf {\bibinfo {volume} {115}},\ \bibinfo {pages} {11150--11155} (\bibinfo {year} {2018})}\BibitemShut {NoStop}%
\bibitem [{\citenamefont {Casci~Ceccacci}\ \emph {et~al.}(2019)\citenamefont {Casci~Ceccacci}, \citenamefont {Cagliani}, \citenamefont {Marizza}, \citenamefont {Schmid},\ and\ \citenamefont {Boisen}}]{casci2019thin}%
  \BibitemOpen
  \bibfield  {author} {\bibinfo {author} {\bibfnamefont {A.}~\bibnamefont {Casci~Ceccacci}}, \bibinfo {author} {\bibfnamefont {A.}~\bibnamefont {Cagliani}}, \bibinfo {author} {\bibfnamefont {P.}~\bibnamefont {Marizza}}, \bibinfo {author} {\bibfnamefont {S.}~\bibnamefont {Schmid}}, \ and\ \bibinfo {author} {\bibfnamefont {A.}~\bibnamefont {Boisen}},\ }\bibfield  {title} {\enquote {\bibinfo {title} {Thin film analysis by nanomechanical infrared spectroscopy},}\ }\href {https://pubs.acs.org/doi/abs/10.1021/acsomega.9b00276} {\bibfield  {journal} {\bibinfo  {journal} {ACS Omega}\ }\textbf {\bibinfo {volume} {4}},\ \bibinfo {pages} {7628--7635} (\bibinfo {year} {2019})}\BibitemShut {NoStop}%
\bibitem [{\citenamefont {Kirchhof}\ \emph {et~al.}(2023)\citenamefont {Kirchhof}, \citenamefont {Yu}, \citenamefont {Yagodkin}, \citenamefont {Stetzuhn}, \citenamefont {de~Ara{\'u}jo}, \citenamefont {Kanellopulos}, \citenamefont {Manas-Valero}, \citenamefont {Coronado}, \citenamefont {van~der Zant}, \citenamefont {Reich} \emph {et~al.}}]{kirchhof2023nanomechanical}%
  \BibitemOpen
  \bibfield  {author} {\bibinfo {author} {\bibfnamefont {J.~N.}\ \bibnamefont {Kirchhof}}, \bibinfo {author} {\bibfnamefont {Y.}~\bibnamefont {Yu}}, \bibinfo {author} {\bibfnamefont {D.}~\bibnamefont {Yagodkin}}, \bibinfo {author} {\bibfnamefont {N.}~\bibnamefont {Stetzuhn}}, \bibinfo {author} {\bibfnamefont {D.~B.}\ \bibnamefont {de~Ara{\'u}jo}}, \bibinfo {author} {\bibfnamefont {K.}~\bibnamefont {Kanellopulos}}, \bibinfo {author} {\bibfnamefont {S.}~\bibnamefont {Manas-Valero}}, \bibinfo {author} {\bibfnamefont {E.}~\bibnamefont {Coronado}}, \bibinfo {author} {\bibfnamefont {H.}~\bibnamefont {van~der Zant}}, \bibinfo {author} {\bibfnamefont {S.}~\bibnamefont {Reich}},  \emph {et~al.},\ }\bibfield  {title} {\enquote {\bibinfo {title} {Nanomechanical absorption spectroscopy of 2d materials with femtowatt sensitivity},}\ }\href {https://iopscience.iop.org/article/10.1088/2053-1583/acd0bf/meta} {\bibfield  {journal} {\bibinfo  {journal} {2D Materials}\ }\textbf {\bibinfo {volume} {10}},\ \bibinfo {pages}
  {035012} (\bibinfo {year} {2023})}\BibitemShut {NoStop}%
\bibitem [{\citenamefont {West}\ \emph {et~al.}(2023)\citenamefont {West}, \citenamefont {Kanellopulos},\ and\ \citenamefont {Schmid}}]{west2023photothermal}%
  \BibitemOpen
  \bibfield  {author} {\bibinfo {author} {\bibfnamefont {R.~G.}\ \bibnamefont {West}}, \bibinfo {author} {\bibfnamefont {K.}~\bibnamefont {Kanellopulos}}, \ and\ \bibinfo {author} {\bibfnamefont {S.}~\bibnamefont {Schmid}},\ }\bibfield  {title} {\enquote {\bibinfo {title} {Photothermal microscopy and spectroscopy with nanomechanical resonators},}\ }\href {https://pubs.acs.org/doi/abs/10.1021/acs.jpcc.3c04361} {\bibfield  {journal} {\bibinfo  {journal} {The Journal of Physical Chemistry C}\ } (\bibinfo {year} {2023})}\BibitemShut {NoStop}%
\bibitem [{\citenamefont {Sementilli}\ \emph {et~al.}(2022)\citenamefont {Sementilli}, \citenamefont {Romero},\ and\ \citenamefont {Bowen}}]{sementilli2022nanomechanical}%
  \BibitemOpen
  \bibfield  {author} {\bibinfo {author} {\bibfnamefont {L.}~\bibnamefont {Sementilli}}, \bibinfo {author} {\bibfnamefont {E.}~\bibnamefont {Romero}}, \ and\ \bibinfo {author} {\bibfnamefont {W.~P.}\ \bibnamefont {Bowen}},\ }\bibfield  {title} {\enquote {\bibinfo {title} {Nanomechanical dissipation and strain engineering},}\ }\href {https://onlinelibrary.wiley.com/doi/abs/10.1002/adfm.202105247} {\bibfield  {journal} {\bibinfo  {journal} {Advanced Functional Materials}\ }\textbf {\bibinfo {volume} {32}},\ \bibinfo {pages} {2105247} (\bibinfo {year} {2022})}\BibitemShut {NoStop}%
\bibitem [{\citenamefont {Usami}\ \emph {et~al.}(2012)\citenamefont {Usami}, \citenamefont {Naesby}, \citenamefont {Bagci}, \citenamefont {Melholt~Nielsen}, \citenamefont {Liu}, \citenamefont {Stobbe}, \citenamefont {Lodahl},\ and\ \citenamefont {Polzik}}]{usami2012optical}%
  \BibitemOpen
  \bibfield  {author} {\bibinfo {author} {\bibfnamefont {K.}~\bibnamefont {Usami}}, \bibinfo {author} {\bibfnamefont {A.}~\bibnamefont {Naesby}}, \bibinfo {author} {\bibfnamefont {T.}~\bibnamefont {Bagci}}, \bibinfo {author} {\bibfnamefont {B.}~\bibnamefont {Melholt~Nielsen}}, \bibinfo {author} {\bibfnamefont {J.}~\bibnamefont {Liu}}, \bibinfo {author} {\bibfnamefont {S.}~\bibnamefont {Stobbe}}, \bibinfo {author} {\bibfnamefont {P.}~\bibnamefont {Lodahl}}, \ and\ \bibinfo {author} {\bibfnamefont {E.~S.}\ \bibnamefont {Polzik}},\ }\bibfield  {title} {\enquote {\bibinfo {title} {Optical cavity cooling of mechanical modes of a semiconductor nanomembrane},}\ }\href {https://www.nature.com/articles/nphys2196} {\bibfield  {journal} {\bibinfo  {journal} {Nature Physics}\ }\textbf {\bibinfo {volume} {8}},\ \bibinfo {pages} {168--172} (\bibinfo {year} {2012})}\BibitemShut {NoStop}%
\bibitem [{\citenamefont {Zhu}\ \emph {et~al.}(2016)\citenamefont {Zhu}, \citenamefont {Yi},\ and\ \citenamefont {Cubukcu}}]{zhu2016plasmonic}%
  \BibitemOpen
  \bibfield  {author} {\bibinfo {author} {\bibfnamefont {H.}~\bibnamefont {Zhu}}, \bibinfo {author} {\bibfnamefont {F.}~\bibnamefont {Yi}}, \ and\ \bibinfo {author} {\bibfnamefont {E.}~\bibnamefont {Cubukcu}},\ }\bibfield  {title} {\enquote {\bibinfo {title} {Plasmonic metamaterial absorber for broadband manipulation of mechanical resonances},}\ }\href {https://www.nature.com/articles/nphoton.2016.183} {\bibfield  {journal} {\bibinfo  {journal} {Nature Photonics}\ }\textbf {\bibinfo {volume} {10}},\ \bibinfo {pages} {709--714} (\bibinfo {year} {2016})}\BibitemShut {NoStop}%
\bibitem [{\citenamefont {Armani}\ and\ \citenamefont {Vahala}(2006)}]{armani2006heavy}%
  \BibitemOpen
  \bibfield  {author} {\bibinfo {author} {\bibfnamefont {A.~M.}\ \bibnamefont {Armani}}\ and\ \bibinfo {author} {\bibfnamefont {K.~J.}\ \bibnamefont {Vahala}},\ }\bibfield  {title} {\enquote {\bibinfo {title} {Heavy water detection using ultra-high-q microcavities},}\ }\href {https://opg.optica.org/abstract.cfm?uri=ol-31-12-1896} {\bibfield  {journal} {\bibinfo  {journal} {Optics Letters}\ }\textbf {\bibinfo {volume} {31}},\ \bibinfo {pages} {1896--1898} (\bibinfo {year} {2006})}\BibitemShut {NoStop}%
\bibitem [{\citenamefont {Nitkowski}\ \emph {et~al.}(2008)\citenamefont {Nitkowski}, \citenamefont {Chen},\ and\ \citenamefont {Lipson}}]{nitkowski2008cavity}%
  \BibitemOpen
  \bibfield  {author} {\bibinfo {author} {\bibfnamefont {A.}~\bibnamefont {Nitkowski}}, \bibinfo {author} {\bibfnamefont {L.}~\bibnamefont {Chen}}, \ and\ \bibinfo {author} {\bibfnamefont {M.}~\bibnamefont {Lipson}},\ }\bibfield  {title} {\enquote {\bibinfo {title} {Cavity-enhanced on-chip absorption spectroscopy using microring resonators},}\ }\href {https://opg.optica.org/abstract.cfm?uri=oe-16-16-11930} {\bibfield  {journal} {\bibinfo  {journal} {Optics Express}\ }\textbf {\bibinfo {volume} {16}},\ \bibinfo {pages} {11930--11936} (\bibinfo {year} {2008})}\BibitemShut {NoStop}%
\bibitem [{\citenamefont {Reinhardt}\ \emph {et~al.}(2016)\citenamefont {Reinhardt}, \citenamefont {M{\"u}ller}, \citenamefont {Bourassa},\ and\ \citenamefont {Sankey}}]{reinhardt2016ultralow}%
  \BibitemOpen
  \bibfield  {author} {\bibinfo {author} {\bibfnamefont {C.}~\bibnamefont {Reinhardt}}, \bibinfo {author} {\bibfnamefont {T.}~\bibnamefont {M{\"u}ller}}, \bibinfo {author} {\bibfnamefont {A.}~\bibnamefont {Bourassa}}, \ and\ \bibinfo {author} {\bibfnamefont {J.~C.}\ \bibnamefont {Sankey}},\ }\bibfield  {title} {\enquote {\bibinfo {title} {Ultralow-noise sin trampoline resonators for sensing and optomechanics},}\ }\href {https://journals.aps.org/prx/abstract/10.1103/PhysRevX.6.021001} {\bibfield  {journal} {\bibinfo  {journal} {Physical Review X}\ }\textbf {\bibinfo {volume} {6}},\ \bibinfo {pages} {021001} (\bibinfo {year} {2016})}\BibitemShut {NoStop}%
\bibitem [{\citenamefont {Pluchar}\ \emph {et~al.}(2020)\citenamefont {Pluchar}, \citenamefont {Agrawal}, \citenamefont {Schenk},\ and\ \citenamefont {Wilson}}]{pluchar2020towards}%
  \BibitemOpen
  \bibfield  {author} {\bibinfo {author} {\bibfnamefont {C.~M.}\ \bibnamefont {Pluchar}}, \bibinfo {author} {\bibfnamefont {A.~R.}\ \bibnamefont {Agrawal}}, \bibinfo {author} {\bibfnamefont {E.}~\bibnamefont {Schenk}}, \ and\ \bibinfo {author} {\bibfnamefont {D.~J.}\ \bibnamefont {Wilson}},\ }\bibfield  {title} {\enquote {\bibinfo {title} {Towards cavity-free ground-state cooling of an acoustic-frequency silicon nitride membrane},}\ }\href {https://opg.optica.org/abstract.cfm?uri=ao-59-22-G107} {\bibfield  {journal} {\bibinfo  {journal} {Applied Optics}\ }\textbf {\bibinfo {volume} {59}},\ \bibinfo {pages} {G107--G111} (\bibinfo {year} {2020})}\BibitemShut {NoStop}%
\bibitem [{\citenamefont {Aspelmeyer}\ \emph {et~al.}(2014)\citenamefont {Aspelmeyer}, \citenamefont {Kippenberg},\ and\ \citenamefont {Marquardt}}]{aspelmeyer2014cavity}%
  \BibitemOpen
  \bibfield  {author} {\bibinfo {author} {\bibfnamefont {M.}~\bibnamefont {Aspelmeyer}}, \bibinfo {author} {\bibfnamefont {T.~J.}\ \bibnamefont {Kippenberg}}, \ and\ \bibinfo {author} {\bibfnamefont {F.}~\bibnamefont {Marquardt}},\ }\bibfield  {title} {\enquote {\bibinfo {title} {Cavity optomechanics},}\ }\href {https://journals.aps.org/rmp/abstract/10.1103/RevModPhys.86.1391} {\bibfield  {journal} {\bibinfo  {journal} {Reviews of Modern Physics}\ }\textbf {\bibinfo {volume} {86}},\ \bibinfo {pages} {1391} (\bibinfo {year} {2014})}\BibitemShut {NoStop}%
\bibitem [{\citenamefont {Zhang}\ \emph {et~al.}(2020)\citenamefont {Zhang}, \citenamefont {Giroux}, \citenamefont {Nour},\ and\ \citenamefont {St-Gelais}}]{zhang2020radiative}%
  \BibitemOpen
  \bibfield  {author} {\bibinfo {author} {\bibfnamefont {C.}~\bibnamefont {Zhang}}, \bibinfo {author} {\bibfnamefont {M.}~\bibnamefont {Giroux}}, \bibinfo {author} {\bibfnamefont {T.~A.}\ \bibnamefont {Nour}}, \ and\ \bibinfo {author} {\bibfnamefont {R.}~\bibnamefont {St-Gelais}},\ }\bibfield  {title} {\enquote {\bibinfo {title} {Radiative heat transfer in freestanding silicon nitride membranes},}\ }\href {https://journals.aps.org/prapplied/abstract/10.1103/PhysRevApplied.14.024072} {\bibfield  {journal} {\bibinfo  {journal} {Physical Review Applied}\ }\textbf {\bibinfo {volume} {14}},\ \bibinfo {pages} {024072} (\bibinfo {year} {2020})}\BibitemShut {NoStop}%
\bibitem [{\citenamefont {Piller}\ \emph {et~al.}(2020)\citenamefont {Piller}, \citenamefont {Sadeghi}, \citenamefont {West}, \citenamefont {Luhmann}, \citenamefont {Martini}, \citenamefont {Hansen},\ and\ \citenamefont {Schmid}}]{piller2020thermal}%
  \BibitemOpen
  \bibfield  {author} {\bibinfo {author} {\bibfnamefont {M.}~\bibnamefont {Piller}}, \bibinfo {author} {\bibfnamefont {P.}~\bibnamefont {Sadeghi}}, \bibinfo {author} {\bibfnamefont {R.~G.}\ \bibnamefont {West}}, \bibinfo {author} {\bibfnamefont {N.}~\bibnamefont {Luhmann}}, \bibinfo {author} {\bibfnamefont {P.}~\bibnamefont {Martini}}, \bibinfo {author} {\bibfnamefont {O.}~\bibnamefont {Hansen}}, \ and\ \bibinfo {author} {\bibfnamefont {S.}~\bibnamefont {Schmid}},\ }\bibfield  {title} {\enquote {\bibinfo {title} {Thermal radiation dominated heat transfer in nanomechanical silicon nitride drum resonators},}\ }\href {https://pubs.aip.org/aip/apl/article-abstract/117/3/034101/312061} {\bibfield  {journal} {\bibinfo  {journal} {Applied Physics Letters}\ }\textbf {\bibinfo {volume} {117}} (\bibinfo {year} {2020})}\BibitemShut {NoStop}%
\bibitem [{\citenamefont {Kla{\ss}}\ \emph {et~al.}(2022)\citenamefont {Kla{\ss}}, \citenamefont {Doster}, \citenamefont {B{\"u}ckle}, \citenamefont {Braive},\ and\ \citenamefont {Weig}}]{klass2022determining}%
  \BibitemOpen
  \bibfield  {author} {\bibinfo {author} {\bibfnamefont {Y.~S.}\ \bibnamefont {Kla{\ss}}}, \bibinfo {author} {\bibfnamefont {J.}~\bibnamefont {Doster}}, \bibinfo {author} {\bibfnamefont {M.}~\bibnamefont {B{\"u}ckle}}, \bibinfo {author} {\bibfnamefont {R.}~\bibnamefont {Braive}}, \ and\ \bibinfo {author} {\bibfnamefont {E.~M.}\ \bibnamefont {Weig}},\ }\bibfield  {title} {\enquote {\bibinfo {title} {Determining young's modulus via the eigenmode spectrum of a nanomechanical string resonator},}\ }\href {https://pubs.aip.org/aip/apl/article/121/8/083501/2834205} {\bibfield  {journal} {\bibinfo  {journal} {Applied Physics Letters}\ }\textbf {\bibinfo {volume} {121}} (\bibinfo {year} {2022})}\BibitemShut {NoStop}%
\bibitem [{\citenamefont {Zhang}\ \emph {et~al.}(2006)\citenamefont {Zhang}, \citenamefont {Krishnaswamy},\ and\ \citenamefont {Lilley}}]{zhang2006bulk}%
  \BibitemOpen
  \bibfield  {author} {\bibinfo {author} {\bibfnamefont {F.}~\bibnamefont {Zhang}}, \bibinfo {author} {\bibfnamefont {S.}~\bibnamefont {Krishnaswamy}}, \ and\ \bibinfo {author} {\bibfnamefont {C.~M.}\ \bibnamefont {Lilley}},\ }\bibfield  {title} {\enquote {\bibinfo {title} {Bulk-wave and guided-wave photoacoustic evaluation of the mechanical properties of aluminum/silicon nitride double-layer thin films},}\ }\href {https://www.sciencedirect.com/science/article/pii/S0041624X06003246?via%3Dihub} {\bibfield  {journal} {\bibinfo  {journal} {Ultrasonics}\ }\textbf {\bibinfo {volume} {45}},\ \bibinfo {pages} {66--76} (\bibinfo {year} {2006})}\BibitemShut {NoStop}%
\bibitem [{\citenamefont {Snell}\ \emph {et~al.}(2022)\citenamefont {Snell}, \citenamefont {Zhang}, \citenamefont {Mu}, \citenamefont {Bouchard},\ and\ \citenamefont {St-Gelais}}]{snell2022heat}%
  \BibitemOpen
  \bibfield  {author} {\bibinfo {author} {\bibfnamefont {N.}~\bibnamefont {Snell}}, \bibinfo {author} {\bibfnamefont {C.}~\bibnamefont {Zhang}}, \bibinfo {author} {\bibfnamefont {G.}~\bibnamefont {Mu}}, \bibinfo {author} {\bibfnamefont {A.}~\bibnamefont {Bouchard}}, \ and\ \bibinfo {author} {\bibfnamefont {R.}~\bibnamefont {St-Gelais}},\ }\bibfield  {title} {\enquote {\bibinfo {title} {Heat transport in silicon nitride drum resonators and its influence on thermal fluctuation-induced frequency noise},}\ }\href {https://journals.aps.org/prapplied/abstract/10.1103/PhysRevApplied.17.044019} {\bibfield  {journal} {\bibinfo  {journal} {Physical Review Applied}\ }\textbf {\bibinfo {volume} {17}},\ \bibinfo {pages} {044019} (\bibinfo {year} {2022})}\BibitemShut {NoStop}%
\bibitem [{\citenamefont {Zhang}\ \emph {et~al.}(2021)\citenamefont {Zhang}, \citenamefont {Bouchard}, \citenamefont {Giroux}, \citenamefont {Nour},\ and\ \citenamefont {St-Gelais}}]{zhang2021erratum}%
  \BibitemOpen
  \bibfield  {author} {\bibinfo {author} {\bibfnamefont {C.}~\bibnamefont {Zhang}}, \bibinfo {author} {\bibfnamefont {A.}~\bibnamefont {Bouchard}}, \bibinfo {author} {\bibfnamefont {M.}~\bibnamefont {Giroux}}, \bibinfo {author} {\bibfnamefont {T.~A.}\ \bibnamefont {Nour}}, \ and\ \bibinfo {author} {\bibfnamefont {R.}~\bibnamefont {St-Gelais}},\ }\bibfield  {title} {\enquote {\bibinfo {title} {Erratum: Radiative heat transfer in freestanding silicon nitride membranes [phys. rev. appl. 14, 024072 (2020)]},}\ }\href {https://journals.aps.org/prapplied/abstract/10.1103/PhysRevApplied.16.019901} {\bibfield  {journal} {\bibinfo  {journal} {Physical Review Applied}\ }\textbf {\bibinfo {volume} {16}},\ \bibinfo {pages} {019901} (\bibinfo {year} {2021})}\BibitemShut {NoStop}%
\bibitem [{\citenamefont {Vivekananthan}\ \emph {et~al.}(2020)\citenamefont {Vivekananthan}, \citenamefont {Ahilan}, \citenamefont {Sakthivelu},\ and\ \citenamefont {Saravanakumar}}]{vivekananthan2020primary}%
  \BibitemOpen
  \bibfield  {author} {\bibinfo {author} {\bibfnamefont {M.}~\bibnamefont {Vivekananthan}}, \bibinfo {author} {\bibfnamefont {C.}~\bibnamefont {Ahilan}}, \bibinfo {author} {\bibfnamefont {S.}~\bibnamefont {Sakthivelu}}, \ and\ \bibinfo {author} {\bibfnamefont {M.}~\bibnamefont {Saravanakumar}},\ }\bibfield  {title} {\enquote {\bibinfo {title} {A primary study of density and compressive strength of the silicon nitride and titanium nitride ceramic composite},}\ }\href {https://www.sciencedirect.com/science/article/pii/S2214785320306854} {\bibfield  {journal} {\bibinfo  {journal} {Materials Today: Proceedings}\ }\textbf {\bibinfo {volume} {33}},\ \bibinfo {pages} {2741--2745} (\bibinfo {year} {2020})}\BibitemShut {NoStop}%
\bibitem [{\citenamefont {Ftouni}\ \emph {et~al.}(2015)\citenamefont {Ftouni}, \citenamefont {Blanc}, \citenamefont {Tainoff}, \citenamefont {Fefferman}, \citenamefont {Defoort}, \citenamefont {Lulla}, \citenamefont {Richard}, \citenamefont {Collin},\ and\ \citenamefont {Bourgeois}}]{ftouni2015thermal}%
  \BibitemOpen
  \bibfield  {author} {\bibinfo {author} {\bibfnamefont {H.}~\bibnamefont {Ftouni}}, \bibinfo {author} {\bibfnamefont {C.}~\bibnamefont {Blanc}}, \bibinfo {author} {\bibfnamefont {D.}~\bibnamefont {Tainoff}}, \bibinfo {author} {\bibfnamefont {A.~D.}\ \bibnamefont {Fefferman}}, \bibinfo {author} {\bibfnamefont {M.}~\bibnamefont {Defoort}}, \bibinfo {author} {\bibfnamefont {K.~J.}\ \bibnamefont {Lulla}}, \bibinfo {author} {\bibfnamefont {J.}~\bibnamefont {Richard}}, \bibinfo {author} {\bibfnamefont {E.}~\bibnamefont {Collin}}, \ and\ \bibinfo {author} {\bibfnamefont {O.}~\bibnamefont {Bourgeois}},\ }\bibfield  {title} {\enquote {\bibinfo {title} {Thermal conductivity of silicon nitride membranes is not sensitive to stress},}\ }\href {https://journals.aps.org/prb/abstract/10.1103/PhysRevB.92.125439} {\bibfield  {journal} {\bibinfo  {journal} {Physical Review B}\ }\textbf {\bibinfo {volume} {92}},\ \bibinfo {pages} {125439} (\bibinfo {year} {2015})}\BibitemShut {NoStop}%
\bibitem [{\citenamefont {Kuwabara}\ \emph {et~al.}(2008)\citenamefont {Kuwabara}, \citenamefont {Matsunaga},\ and\ \citenamefont {Tanaka}}]{kuwabara2008lattice}%
  \BibitemOpen
  \bibfield  {author} {\bibinfo {author} {\bibfnamefont {A.}~\bibnamefont {Kuwabara}}, \bibinfo {author} {\bibfnamefont {K.}~\bibnamefont {Matsunaga}}, \ and\ \bibinfo {author} {\bibfnamefont {I.}~\bibnamefont {Tanaka}},\ }\bibfield  {title} {\enquote {\bibinfo {title} {Lattice dynamics and thermodynamical properties of silicon nitride polymorphs},}\ }\href {https://journals.aps.org/prb/abstract/10.1103/PhysRevB.78.064104} {\bibfield  {journal} {\bibinfo  {journal} {Physical Review B}\ }\textbf {\bibinfo {volume} {78}},\ \bibinfo {pages} {064104} (\bibinfo {year} {2008})}\BibitemShut {NoStop}%
\bibitem [{\citenamefont {Philipp}(1973)}]{philipp1973optical}%
  \BibitemOpen
  \bibfield  {author} {\bibinfo {author} {\bibfnamefont {H.~R.}\ \bibnamefont {Philipp}},\ }\bibfield  {title} {\enquote {\bibinfo {title} {Optical properties of silicon nitride},}\ }\href {https://iopscience.iop.org/article/10.1149/1.2403440} {\bibfield  {journal} {\bibinfo  {journal} {Journal of the Electrochemical Society}\ }\textbf {\bibinfo {volume} {120}},\ \bibinfo {pages} {295} (\bibinfo {year} {1973})}\BibitemShut {NoStop}%
\bibitem [{\citenamefont {Bauters}\ \emph {et~al.}(2011)\citenamefont {Bauters}, \citenamefont {Heck}, \citenamefont {John}, \citenamefont {Dai}, \citenamefont {Tien}, \citenamefont {Barton}, \citenamefont {Leinse}, \citenamefont {Heideman}, \citenamefont {Blumenthal},\ and\ \citenamefont {Bowers}}]{bauters2011ultra}%
  \BibitemOpen
  \bibfield  {author} {\bibinfo {author} {\bibfnamefont {J.~F.}\ \bibnamefont {Bauters}}, \bibinfo {author} {\bibfnamefont {M.~J.}\ \bibnamefont {Heck}}, \bibinfo {author} {\bibfnamefont {D.}~\bibnamefont {John}}, \bibinfo {author} {\bibfnamefont {D.}~\bibnamefont {Dai}}, \bibinfo {author} {\bibfnamefont {M.-C.}\ \bibnamefont {Tien}}, \bibinfo {author} {\bibfnamefont {J.~S.}\ \bibnamefont {Barton}}, \bibinfo {author} {\bibfnamefont {A.}~\bibnamefont {Leinse}}, \bibinfo {author} {\bibfnamefont {R.~G.}\ \bibnamefont {Heideman}}, \bibinfo {author} {\bibfnamefont {D.~J.}\ \bibnamefont {Blumenthal}}, \ and\ \bibinfo {author} {\bibfnamefont {J.~E.}\ \bibnamefont {Bowers}},\ }\bibfield  {title} {\enquote {\bibinfo {title} {Ultra-low-loss high-aspect-ratio si 3 n 4 waveguides},}\ }\href {https://opg.optica.org/oe/fulltext.cfm?uri=oe-19-4-3163&id=209915} {\bibfield  {journal} {\bibinfo  {journal} {Optics Express}\ }\textbf {\bibinfo {volume} {19}},\ \bibinfo {pages} {3163--3174} (\bibinfo {year} {2011})}\BibitemShut
  {NoStop}%
\bibitem [{\citenamefont {Xiong}\ \emph {et~al.}(2020)\citenamefont {Xiong}, \citenamefont {Jiang}, \citenamefont {Li}, \citenamefont {Zhang}, \citenamefont {Xu}, \citenamefont {Zhao}, \citenamefont {Wang}, \citenamefont {Liu}, \citenamefont {Luo}, \citenamefont {Li} \emph {et~al.}}]{xiong2020sin}%
  \BibitemOpen
  \bibfield  {author} {\bibinfo {author} {\bibfnamefont {W.}~\bibnamefont {Xiong}}, \bibinfo {author} {\bibfnamefont {H.}~\bibnamefont {Jiang}}, \bibinfo {author} {\bibfnamefont {T.}~\bibnamefont {Li}}, \bibinfo {author} {\bibfnamefont {P.}~\bibnamefont {Zhang}}, \bibinfo {author} {\bibfnamefont {Q.}~\bibnamefont {Xu}}, \bibinfo {author} {\bibfnamefont {X.}~\bibnamefont {Zhao}}, \bibinfo {author} {\bibfnamefont {G.}~\bibnamefont {Wang}}, \bibinfo {author} {\bibfnamefont {Y.}~\bibnamefont {Liu}}, \bibinfo {author} {\bibfnamefont {Y.}~\bibnamefont {Luo}}, \bibinfo {author} {\bibfnamefont {Z.}~\bibnamefont {Li}},  \emph {et~al.},\ }\bibfield  {title} {\enquote {\bibinfo {title} {Sin x films and membranes for photonic and mems applications},}\ }\href {https://link.springer.com/article/10.1007/s10854-019-01164-9} {\bibfield  {journal} {\bibinfo  {journal} {Journal of Materials Science: Materials in Electronics}\ }\textbf {\bibinfo {volume} {31}},\ \bibinfo {pages} {90--97} (\bibinfo {year} {2020})}\BibitemShut
  {NoStop}%
\bibitem [{\citenamefont {Wang}\ \emph {et~al.}(2018)\citenamefont {Wang}, \citenamefont {Xie}, \citenamefont {Van~Thourhout}, \citenamefont {Zhang}, \citenamefont {Yu},\ and\ \citenamefont {Wang}}]{wang2018nonlinear}%
  \BibitemOpen
  \bibfield  {author} {\bibinfo {author} {\bibfnamefont {L.}~\bibnamefont {Wang}}, \bibinfo {author} {\bibfnamefont {W.}~\bibnamefont {Xie}}, \bibinfo {author} {\bibfnamefont {D.}~\bibnamefont {Van~Thourhout}}, \bibinfo {author} {\bibfnamefont {Y.}~\bibnamefont {Zhang}}, \bibinfo {author} {\bibfnamefont {H.}~\bibnamefont {Yu}}, \ and\ \bibinfo {author} {\bibfnamefont {S.}~\bibnamefont {Wang}},\ }\bibfield  {title} {\enquote {\bibinfo {title} {Nonlinear silicon nitride waveguides based on a pecvd deposition platform},}\ }\href {https://opg.optica.org/oe/fulltext.cfm?uri=oe-26-8-9645&id=385310} {\bibfield  {journal} {\bibinfo  {journal} {Optics Express}\ }\textbf {\bibinfo {volume} {26}},\ \bibinfo {pages} {9645--9654} (\bibinfo {year} {2018})}\BibitemShut {NoStop}%
\bibitem [{\citenamefont {Daldosso}\ \emph {et~al.}(2004{\natexlab{a}})\citenamefont {Daldosso}, \citenamefont {Melchiorri}, \citenamefont {Riboli}, \citenamefont {Girardini}, \citenamefont {Pucker}, \citenamefont {Crivellari}, \citenamefont {Bellutti}, \citenamefont {Lui},\ and\ \citenamefont {Pavesi}}]{daldosso2004comparison}%
  \BibitemOpen
  \bibfield  {author} {\bibinfo {author} {\bibfnamefont {N.}~\bibnamefont {Daldosso}}, \bibinfo {author} {\bibfnamefont {M.}~\bibnamefont {Melchiorri}}, \bibinfo {author} {\bibfnamefont {F.}~\bibnamefont {Riboli}}, \bibinfo {author} {\bibfnamefont {M.}~\bibnamefont {Girardini}}, \bibinfo {author} {\bibfnamefont {G.}~\bibnamefont {Pucker}}, \bibinfo {author} {\bibfnamefont {M.}~\bibnamefont {Crivellari}}, \bibinfo {author} {\bibfnamefont {P.}~\bibnamefont {Bellutti}}, \bibinfo {author} {\bibfnamefont {A.}~\bibnamefont {Lui}}, \ and\ \bibinfo {author} {\bibfnamefont {L.}~\bibnamefont {Pavesi}},\ }\bibfield  {title} {\enquote {\bibinfo {title} {Comparison among various si/sub 3/n/sub 4/waveguide geometries grown within a cmos fabrication pilot line},}\ }\href {https://ieeexplore.ieee.org/abstract/document/1310422} {\bibfield  {journal} {\bibinfo  {journal} {Journal of Lightwave Technology}\ }\textbf {\bibinfo {volume} {22}},\ \bibinfo {pages} {1734--1740} (\bibinfo {year} {2004}{\natexlab{a}})}\BibitemShut
  {NoStop}%
\bibitem [{\citenamefont {Bulla}\ \emph {et~al.}(1999)\citenamefont {Bulla}, \citenamefont {Borges}, \citenamefont {Romero}, \citenamefont {Morimoto}, \citenamefont {Neto},\ and\ \citenamefont {Cortes}}]{bulla1999design}%
  \BibitemOpen
  \bibfield  {author} {\bibinfo {author} {\bibfnamefont {D.}~\bibnamefont {Bulla}}, \bibinfo {author} {\bibfnamefont {B.}~\bibnamefont {Borges}}, \bibinfo {author} {\bibfnamefont {M.}~\bibnamefont {Romero}}, \bibinfo {author} {\bibfnamefont {N.}~\bibnamefont {Morimoto}}, \bibinfo {author} {\bibfnamefont {L.}~\bibnamefont {Neto}}, \ and\ \bibinfo {author} {\bibfnamefont {A.}~\bibnamefont {Cortes}},\ }\bibfield  {title} {\enquote {\bibinfo {title} {Design and fabrication of sio/sub 2//si/sub 3/n/sub 4/cvd optical waveguides},}\ }in\ \href {https://ieeexplore.ieee.org/abstract/document/866156} {\emph {\bibinfo {booktitle} {1999 SBMO/IEEE MTT-S International Microwave and Optoelectronics Conference}}},\ Vol.~\bibinfo {volume} {2}\ (\bibinfo {organization} {IEEE},\ \bibinfo {year} {1999})\ pp.\ \bibinfo {pages} {454--457}\BibitemShut {NoStop}%
\bibitem [{\citenamefont {Inukai}\ and\ \citenamefont {Ono}(1994)}]{inukai1994optical}%
  \BibitemOpen
  \bibfield  {author} {\bibinfo {author} {\bibfnamefont {T.~I.~T.}\ \bibnamefont {Inukai}}\ and\ \bibinfo {author} {\bibfnamefont {K.~O.~K.}\ \bibnamefont {Ono}},\ }\bibfield  {title} {\enquote {\bibinfo {title} {Optical characteristics of amorphous silicon nitride thin films prepared by electron cyclotron resonance plasma chemical vapor deposition},}\ }\href {https://iopscience.iop.org/article/10.1143/JJAP.33.2593} {\bibfield  {journal} {\bibinfo  {journal} {Japanese Journal of Applied Physics}\ }\textbf {\bibinfo {volume} {33}},\ \bibinfo {pages} {2593} (\bibinfo {year} {1994})}\BibitemShut {NoStop}%
\bibitem [{\citenamefont {Daldosso}\ \emph {et~al.}(2004{\natexlab{b}})\citenamefont {Daldosso}, \citenamefont {Melchiorri}, \citenamefont {Riboli}, \citenamefont {Sbrana}, \citenamefont {Pavesi}, \citenamefont {Pucker}, \citenamefont {Kompocholis}, \citenamefont {Crivellari}, \citenamefont {Bellutti},\ and\ \citenamefont {Lui}}]{daldosso2004fabrication}%
  \BibitemOpen
  \bibfield  {author} {\bibinfo {author} {\bibfnamefont {N.}~\bibnamefont {Daldosso}}, \bibinfo {author} {\bibfnamefont {M.}~\bibnamefont {Melchiorri}}, \bibinfo {author} {\bibfnamefont {F.}~\bibnamefont {Riboli}}, \bibinfo {author} {\bibfnamefont {F.}~\bibnamefont {Sbrana}}, \bibinfo {author} {\bibfnamefont {L.}~\bibnamefont {Pavesi}}, \bibinfo {author} {\bibfnamefont {G.}~\bibnamefont {Pucker}}, \bibinfo {author} {\bibfnamefont {C.}~\bibnamefont {Kompocholis}}, \bibinfo {author} {\bibfnamefont {M.}~\bibnamefont {Crivellari}}, \bibinfo {author} {\bibfnamefont {P.}~\bibnamefont {Bellutti}}, \ and\ \bibinfo {author} {\bibfnamefont {A.}~\bibnamefont {Lui}},\ }\bibfield  {title} {\enquote {\bibinfo {title} {Fabrication and optical characterization of thin two-dimensional si3n4 waveguides},}\ }\href {https://www.sciencedirect.com/science/article/pii/S1369800104000757} {\bibfield  {journal} {\bibinfo  {journal} {Materials Science in Semiconductor Processing}\ }\textbf {\bibinfo {volume} {7}},\ \bibinfo {pages}
  {453--458} (\bibinfo {year} {2004}{\natexlab{b}})}\BibitemShut {NoStop}%
\bibitem [{\citenamefont {Buzaverov}\ \emph {et~al.}(2023)\citenamefont {Buzaverov}, \citenamefont {Baburin}, \citenamefont {Sergeev}, \citenamefont {Avdeev}, \citenamefont {Lotkov}, \citenamefont {Andronik}, \citenamefont {Stukalova}, \citenamefont {Baklykov}, \citenamefont {Dyakonov}, \citenamefont {Skryabin} \emph {et~al.}}]{buzaverov2023low}%
  \BibitemOpen
  \bibfield  {author} {\bibinfo {author} {\bibfnamefont {K.~A.}\ \bibnamefont {Buzaverov}}, \bibinfo {author} {\bibfnamefont {A.~S.}\ \bibnamefont {Baburin}}, \bibinfo {author} {\bibfnamefont {E.~V.}\ \bibnamefont {Sergeev}}, \bibinfo {author} {\bibfnamefont {S.~S.}\ \bibnamefont {Avdeev}}, \bibinfo {author} {\bibfnamefont {E.~S.}\ \bibnamefont {Lotkov}}, \bibinfo {author} {\bibfnamefont {M.}~\bibnamefont {Andronik}}, \bibinfo {author} {\bibfnamefont {V.~E.}\ \bibnamefont {Stukalova}}, \bibinfo {author} {\bibfnamefont {D.~A.}\ \bibnamefont {Baklykov}}, \bibinfo {author} {\bibfnamefont {I.~V.}\ \bibnamefont {Dyakonov}}, \bibinfo {author} {\bibfnamefont {N.~N.}\ \bibnamefont {Skryabin}},  \emph {et~al.},\ }\bibfield  {title} {\enquote {\bibinfo {title} {Low-loss silicon nitride photonic ics for near-infrared wavelength bandwidth},}\ }\href {https://opg.optica.org/oe/fulltext.cfm?uri=oe-31-10-16227&id=530269} {\bibfield  {journal} {\bibinfo  {journal} {Optics Express}\ }\textbf {\bibinfo {volume} {31}},\
  \bibinfo {pages} {16227--16242} (\bibinfo {year} {2023})}\BibitemShut {NoStop}%
\bibitem [{\citenamefont {Smith}\ \emph {et~al.}(2023)\citenamefont {Smith}, \citenamefont {Francis}, \citenamefont {Navickaite},\ and\ \citenamefont {Strain}}]{smith2023sin}%
  \BibitemOpen
  \bibfield  {author} {\bibinfo {author} {\bibfnamefont {J.~A.}\ \bibnamefont {Smith}}, \bibinfo {author} {\bibfnamefont {H.}~\bibnamefont {Francis}}, \bibinfo {author} {\bibfnamefont {G.}~\bibnamefont {Navickaite}}, \ and\ \bibinfo {author} {\bibfnamefont {M.~J.}\ \bibnamefont {Strain}},\ }\bibfield  {title} {\enquote {\bibinfo {title} {Sin foundry platform for high performance visible light integrated photonics},}\ }\href {https://opg.optica.org/ome/fulltext.cfm?uri=ome-13-2-458&id=525470} {\bibfield  {journal} {\bibinfo  {journal} {Optical Materials Express}\ }\textbf {\bibinfo {volume} {13}},\ \bibinfo {pages} {458--468} (\bibinfo {year} {2023})}\BibitemShut {NoStop}%
\bibitem [{\citenamefont {Yong}\ \emph {et~al.}(2022)\citenamefont {Yong}, \citenamefont {Chen}, \citenamefont {Luo}, \citenamefont {Govdeli}, \citenamefont {Chua}, \citenamefont {Azadeh}, \citenamefont {Stalmashonak}, \citenamefont {Lo}, \citenamefont {Poon},\ and\ \citenamefont {Sacher}}]{yong2022power}%
  \BibitemOpen
  \bibfield  {author} {\bibinfo {author} {\bibfnamefont {Z.}~\bibnamefont {Yong}}, \bibinfo {author} {\bibfnamefont {H.}~\bibnamefont {Chen}}, \bibinfo {author} {\bibfnamefont {X.}~\bibnamefont {Luo}}, \bibinfo {author} {\bibfnamefont {A.}~\bibnamefont {Govdeli}}, \bibinfo {author} {\bibfnamefont {H.}~\bibnamefont {Chua}}, \bibinfo {author} {\bibfnamefont {S.~S.}\ \bibnamefont {Azadeh}}, \bibinfo {author} {\bibfnamefont {A.}~\bibnamefont {Stalmashonak}}, \bibinfo {author} {\bibfnamefont {G.-Q.}\ \bibnamefont {Lo}}, \bibinfo {author} {\bibfnamefont {J.~K.}\ \bibnamefont {Poon}}, \ and\ \bibinfo {author} {\bibfnamefont {W.~D.}\ \bibnamefont {Sacher}},\ }\bibfield  {title} {\enquote {\bibinfo {title} {Power-efficient silicon nitride thermo-optic phase shifters for visible light},}\ }\href {https://opg.optica.org/oe/fulltext.cfm?uri=oe-30-5-7225&id=469588} {\bibfield  {journal} {\bibinfo  {journal} {Optics Express}\ }\textbf {\bibinfo {volume} {30}},\ \bibinfo {pages} {7225--7237} (\bibinfo {year}
  {2022})}\BibitemShut {NoStop}%
\bibitem [{\citenamefont {Sacher}\ \emph {et~al.}(2019)\citenamefont {Sacher}, \citenamefont {Luo}, \citenamefont {Yang}, \citenamefont {Chen}, \citenamefont {Lordello}, \citenamefont {Mak}, \citenamefont {Liu}, \citenamefont {Hu}, \citenamefont {Xue}, \citenamefont {Lo} \emph {et~al.}}]{sacher2019visible}%
  \BibitemOpen
  \bibfield  {author} {\bibinfo {author} {\bibfnamefont {W.~D.}\ \bibnamefont {Sacher}}, \bibinfo {author} {\bibfnamefont {X.}~\bibnamefont {Luo}}, \bibinfo {author} {\bibfnamefont {Y.}~\bibnamefont {Yang}}, \bibinfo {author} {\bibfnamefont {F.-D.}\ \bibnamefont {Chen}}, \bibinfo {author} {\bibfnamefont {T.}~\bibnamefont {Lordello}}, \bibinfo {author} {\bibfnamefont {J.~C.}\ \bibnamefont {Mak}}, \bibinfo {author} {\bibfnamefont {X.}~\bibnamefont {Liu}}, \bibinfo {author} {\bibfnamefont {T.}~\bibnamefont {Hu}}, \bibinfo {author} {\bibfnamefont {T.}~\bibnamefont {Xue}}, \bibinfo {author} {\bibfnamefont {P.~G.-Q.}\ \bibnamefont {Lo}},  \emph {et~al.},\ }\bibfield  {title} {\enquote {\bibinfo {title} {Visible-light silicon nitride waveguide devices and implantable neurophotonic probes on thinned 200 mm silicon wafers},}\ }\href {https://opg.optica.org/oe/fulltext.cfm?uri=oe-27-26-37400&id=424028} {\bibfield  {journal} {\bibinfo  {journal} {Optics Express}\ }\textbf {\bibinfo {volume} {27}},\ \bibinfo {pages}
  {37400--37418} (\bibinfo {year} {2019})}\BibitemShut {NoStop}%
\bibitem [{\citenamefont {Subramanian}\ \emph {et~al.}(2013)\citenamefont {Subramanian}, \citenamefont {Neutens}, \citenamefont {Dhakal}, \citenamefont {Jansen}, \citenamefont {Claes}, \citenamefont {Rottenberg}, \citenamefont {Peyskens}, \citenamefont {Selvaraja}, \citenamefont {Helin}, \citenamefont {Du~Bois} \emph {et~al.}}]{subramanian2013low}%
  \BibitemOpen
  \bibfield  {author} {\bibinfo {author} {\bibfnamefont {A.}~\bibnamefont {Subramanian}}, \bibinfo {author} {\bibfnamefont {P.}~\bibnamefont {Neutens}}, \bibinfo {author} {\bibfnamefont {A.}~\bibnamefont {Dhakal}}, \bibinfo {author} {\bibfnamefont {R.}~\bibnamefont {Jansen}}, \bibinfo {author} {\bibfnamefont {T.}~\bibnamefont {Claes}}, \bibinfo {author} {\bibfnamefont {X.}~\bibnamefont {Rottenberg}}, \bibinfo {author} {\bibfnamefont {F.}~\bibnamefont {Peyskens}}, \bibinfo {author} {\bibfnamefont {S.}~\bibnamefont {Selvaraja}}, \bibinfo {author} {\bibfnamefont {P.}~\bibnamefont {Helin}}, \bibinfo {author} {\bibfnamefont {B.}~\bibnamefont {Du~Bois}},  \emph {et~al.},\ }\bibfield  {title} {\enquote {\bibinfo {title} {Low-loss singlemode pecvd silicon nitride photonic wire waveguides for 532--900 nm wavelength window fabricated within a cmos pilot line},}\ }\href {https://ieeexplore.ieee.org/document/6674990} {\bibfield  {journal} {\bibinfo  {journal} {IEEE Photonics Journal}\ }\textbf {\bibinfo {volume} {5}},\
  \bibinfo {pages} {2202809--2202809} (\bibinfo {year} {2013})}\BibitemShut {NoStop}%
\bibitem [{\citenamefont {Liu}\ \emph {et~al.}(2021)\citenamefont {Liu}, \citenamefont {Huang}, \citenamefont {Wang}, \citenamefont {He}, \citenamefont {Raja}, \citenamefont {Liu}, \citenamefont {Engelsen},\ and\ \citenamefont {Kippenberg}}]{liu2021high}%
  \BibitemOpen
  \bibfield  {author} {\bibinfo {author} {\bibfnamefont {J.}~\bibnamefont {Liu}}, \bibinfo {author} {\bibfnamefont {G.}~\bibnamefont {Huang}}, \bibinfo {author} {\bibfnamefont {R.~N.}\ \bibnamefont {Wang}}, \bibinfo {author} {\bibfnamefont {J.}~\bibnamefont {He}}, \bibinfo {author} {\bibfnamefont {A.~S.}\ \bibnamefont {Raja}}, \bibinfo {author} {\bibfnamefont {T.}~\bibnamefont {Liu}}, \bibinfo {author} {\bibfnamefont {N.~J.}\ \bibnamefont {Engelsen}}, \ and\ \bibinfo {author} {\bibfnamefont {T.~J.}\ \bibnamefont {Kippenberg}},\ }\bibfield  {title} {\enquote {\bibinfo {title} {High-yield, wafer-scale fabrication of ultralow-loss, dispersion-engineered silicon nitride photonic circuits},}\ }\href {https://www.nature.com/articles/s41467-021-21973-z} {\bibfield  {journal} {\bibinfo  {journal} {Nature Communications}\ }\textbf {\bibinfo {volume} {12}},\ \bibinfo {pages} {2236} (\bibinfo {year} {2021})}\BibitemShut {NoStop}%
\bibitem [{\citenamefont {Ji}\ \emph {et~al.}(2023)\citenamefont {Ji}, \citenamefont {Okawachi}, \citenamefont {Gil-Molina}, \citenamefont {Corato-Zanarella}, \citenamefont {Roberts}, \citenamefont {Gaeta},\ and\ \citenamefont {Lipson}}]{ji2023ultra}%
  \BibitemOpen
  \bibfield  {author} {\bibinfo {author} {\bibfnamefont {X.}~\bibnamefont {Ji}}, \bibinfo {author} {\bibfnamefont {Y.}~\bibnamefont {Okawachi}}, \bibinfo {author} {\bibfnamefont {A.}~\bibnamefont {Gil-Molina}}, \bibinfo {author} {\bibfnamefont {M.}~\bibnamefont {Corato-Zanarella}}, \bibinfo {author} {\bibfnamefont {S.}~\bibnamefont {Roberts}}, \bibinfo {author} {\bibfnamefont {A.~L.}\ \bibnamefont {Gaeta}}, \ and\ \bibinfo {author} {\bibfnamefont {M.}~\bibnamefont {Lipson}},\ }\bibfield  {title} {\enquote {\bibinfo {title} {Ultra-low-loss silicon nitride photonics based on deposited films compatible with foundries},}\ }\href {https://onlinelibrary.wiley.com/doi/full/10.1002/lpor.202200544} {\bibfield  {journal} {\bibinfo  {journal} {Laser \& Photonics Reviews}\ }\textbf {\bibinfo {volume} {17}},\ \bibinfo {pages} {2200544} (\bibinfo {year} {2023})}\BibitemShut {NoStop}%
\bibitem [{\citenamefont {Zhang}\ \emph {et~al.}(2023)\citenamefont {Zhang}, \citenamefont {Bi}, \citenamefont {Harder}, \citenamefont {Lohse}, \citenamefont {Gannott}, \citenamefont {Gumann}, \citenamefont {Zhang},\ and\ \citenamefont {Del'Haye}}]{zhang2023room}%
  \BibitemOpen
  \bibfield  {author} {\bibinfo {author} {\bibfnamefont {S.}~\bibnamefont {Zhang}}, \bibinfo {author} {\bibfnamefont {T.}~\bibnamefont {Bi}}, \bibinfo {author} {\bibfnamefont {I.}~\bibnamefont {Harder}}, \bibinfo {author} {\bibfnamefont {O.}~\bibnamefont {Lohse}}, \bibinfo {author} {\bibfnamefont {F.}~\bibnamefont {Gannott}}, \bibinfo {author} {\bibfnamefont {A.}~\bibnamefont {Gumann}}, \bibinfo {author} {\bibfnamefont {Y.}~\bibnamefont {Zhang}}, \ and\ \bibinfo {author} {\bibfnamefont {P.}~\bibnamefont {Del'Haye}},\ }\bibfield  {title} {\enquote {\bibinfo {title} {Room-temperature sputtered ultralow-loss silicon nitride},}\ }in\ \href {https://ieeexplore.ieee.org/abstract/document/10232644} {\emph {\bibinfo {booktitle} {2023 Conference on Lasers and Electro-Optics Europe \& European Quantum Electronics Conference (CLEO/Europe-EQEC)}}}\ (\bibinfo {organization} {IEEE},\ \bibinfo {year} {2023})\ pp.\ \bibinfo {pages} {1--1}\BibitemShut {NoStop}%
\bibitem [{\citenamefont {Karuza}\ \emph {et~al.}(2012)\citenamefont {Karuza}, \citenamefont {Galassi}, \citenamefont {Biancofiore}, \citenamefont {Molinelli}, \citenamefont {Natali}, \citenamefont {Tombesi}, \citenamefont {Di~Giuseppe},\ and\ \citenamefont {Vitali}}]{karuza2012tunable}%
  \BibitemOpen
  \bibfield  {author} {\bibinfo {author} {\bibfnamefont {M.}~\bibnamefont {Karuza}}, \bibinfo {author} {\bibfnamefont {M.}~\bibnamefont {Galassi}}, \bibinfo {author} {\bibfnamefont {C.}~\bibnamefont {Biancofiore}}, \bibinfo {author} {\bibfnamefont {C.}~\bibnamefont {Molinelli}}, \bibinfo {author} {\bibfnamefont {R.}~\bibnamefont {Natali}}, \bibinfo {author} {\bibfnamefont {P.}~\bibnamefont {Tombesi}}, \bibinfo {author} {\bibfnamefont {G.}~\bibnamefont {Di~Giuseppe}}, \ and\ \bibinfo {author} {\bibfnamefont {D.}~\bibnamefont {Vitali}},\ }\bibfield  {title} {\enquote {\bibinfo {title} {Tunable linear and quadratic optomechanical coupling for a tilted membrane within an optical cavity: theory and experiment},}\ }\href {https://iopscience.iop.org/article/10.1088/2040-8978/15/2/025704} {\bibfield  {journal} {\bibinfo  {journal} {Journal of Optics}\ }\textbf {\bibinfo {volume} {15}},\ \bibinfo {pages} {025704} (\bibinfo {year} {2012})}\BibitemShut {NoStop}%
\bibitem [{\citenamefont {Wilson}\ \emph {et~al.}(2009)\citenamefont {Wilson}, \citenamefont {Regal}, \citenamefont {Papp},\ and\ \citenamefont {Kimble}}]{wilson2009cavity}%
  \BibitemOpen
  \bibfield  {author} {\bibinfo {author} {\bibfnamefont {D.~J.}\ \bibnamefont {Wilson}}, \bibinfo {author} {\bibfnamefont {C.~A.}\ \bibnamefont {Regal}}, \bibinfo {author} {\bibfnamefont {S.~B.}\ \bibnamefont {Papp}}, \ and\ \bibinfo {author} {\bibfnamefont {H.}~\bibnamefont {Kimble}},\ }\bibfield  {title} {\enquote {\bibinfo {title} {Cavity optomechanics with stoichiometric sin films},}\ }\href {https://journals.aps.org/prl/abstract/10.1103/PhysRevLett.103.207204} {\bibfield  {journal} {\bibinfo  {journal} {Physical Review Letters}\ }\textbf {\bibinfo {volume} {103}},\ \bibinfo {pages} {207204} (\bibinfo {year} {2009})}\BibitemShut {NoStop}%
\bibitem [{\citenamefont {Wilson}(2012)}]{wilson2012cavity}%
  \BibitemOpen
  \bibfield  {author} {\bibinfo {author} {\bibfnamefont {D.~J.}\ \bibnamefont {Wilson}},\ }\href {https://thesis.library.caltech.edu/7162/} {\emph {\bibinfo {title} {Cavity optomechanics with high-stress silicon nitride films}}}\ (\bibinfo  {publisher} {California Institute of Technology},\ \bibinfo {year} {2012})\BibitemShut {NoStop}%
\bibitem [{\citenamefont {Serra}\ \emph {et~al.}(2016)\citenamefont {Serra}, \citenamefont {Bawaj}, \citenamefont {Borrielli}, \citenamefont {Di~Giuseppe}, \citenamefont {Forte}, \citenamefont {Kralj}, \citenamefont {Malossi}, \citenamefont {Marconi}, \citenamefont {Marin}, \citenamefont {Marino} \emph {et~al.}}]{serra2016microfabrication}%
  \BibitemOpen
  \bibfield  {author} {\bibinfo {author} {\bibfnamefont {E.}~\bibnamefont {Serra}}, \bibinfo {author} {\bibfnamefont {M.}~\bibnamefont {Bawaj}}, \bibinfo {author} {\bibfnamefont {A.}~\bibnamefont {Borrielli}}, \bibinfo {author} {\bibfnamefont {G.}~\bibnamefont {Di~Giuseppe}}, \bibinfo {author} {\bibfnamefont {S.}~\bibnamefont {Forte}}, \bibinfo {author} {\bibfnamefont {N.}~\bibnamefont {Kralj}}, \bibinfo {author} {\bibfnamefont {N.}~\bibnamefont {Malossi}}, \bibinfo {author} {\bibfnamefont {L.}~\bibnamefont {Marconi}}, \bibinfo {author} {\bibfnamefont {F.}~\bibnamefont {Marin}}, \bibinfo {author} {\bibfnamefont {F.}~\bibnamefont {Marino}},  \emph {et~al.},\ }\bibfield  {title} {\enquote {\bibinfo {title} {Microfabrication of large-area circular high-stress silicon nitride membranes for optomechanical applications},}\ }\href {https://pubs.aip.org/aip/adv/article/6/6/065004/22559} {\bibfield  {journal} {\bibinfo  {journal} {AIP Advances}\ }\textbf {\bibinfo {volume} {6}} (\bibinfo {year} {2016})}\BibitemShut
  {NoStop}%
\bibitem [{\citenamefont {Sankey}\ \emph {et~al.}(2010)\citenamefont {Sankey}, \citenamefont {Yang}, \citenamefont {Zwickl}, \citenamefont {Jayich},\ and\ \citenamefont {Harris}}]{sankey2010strong}%
  \BibitemOpen
  \bibfield  {author} {\bibinfo {author} {\bibfnamefont {J.~C.}\ \bibnamefont {Sankey}}, \bibinfo {author} {\bibfnamefont {C.}~\bibnamefont {Yang}}, \bibinfo {author} {\bibfnamefont {B.~M.}\ \bibnamefont {Zwickl}}, \bibinfo {author} {\bibfnamefont {A.~M.}\ \bibnamefont {Jayich}}, \ and\ \bibinfo {author} {\bibfnamefont {J.~G.}\ \bibnamefont {Harris}},\ }\bibfield  {title} {\enquote {\bibinfo {title} {Strong and tunable nonlinear optomechanical coupling in a low-loss system},}\ }\href {https://www.nature.com/articles/nphys1707} {\bibfield  {journal} {\bibinfo  {journal} {Nature Physics}\ }\textbf {\bibinfo {volume} {6}},\ \bibinfo {pages} {707--712} (\bibinfo {year} {2010})}\BibitemShut {NoStop}%
\bibitem [{\citenamefont {Weaver}\ \emph {et~al.}(2017)\citenamefont {Weaver}, \citenamefont {Buters}, \citenamefont {Luna}, \citenamefont {Eerkens}, \citenamefont {Heeck}, \citenamefont {de~Man},\ and\ \citenamefont {Bouwmeester}}]{weaver2017coherent}%
  \BibitemOpen
  \bibfield  {author} {\bibinfo {author} {\bibfnamefont {M.~J.}\ \bibnamefont {Weaver}}, \bibinfo {author} {\bibfnamefont {F.}~\bibnamefont {Buters}}, \bibinfo {author} {\bibfnamefont {F.}~\bibnamefont {Luna}}, \bibinfo {author} {\bibfnamefont {H.}~\bibnamefont {Eerkens}}, \bibinfo {author} {\bibfnamefont {K.}~\bibnamefont {Heeck}}, \bibinfo {author} {\bibfnamefont {S.}~\bibnamefont {de~Man}}, \ and\ \bibinfo {author} {\bibfnamefont {D.}~\bibnamefont {Bouwmeester}},\ }\bibfield  {title} {\enquote {\bibinfo {title} {Coherent optomechanical state transfer between disparate mechanical resonators},}\ }\href {https://www.nature.com/articles/s41467-017-00968-9} {\bibfield  {journal} {\bibinfo  {journal} {Nature Communications}\ }\textbf {\bibinfo {volume} {8}},\ \bibinfo {pages} {824} (\bibinfo {year} {2017})}\BibitemShut {NoStop}%
\bibitem [{\citenamefont {Stambaugh}\ \emph {et~al.}(2014)\citenamefont {Stambaugh}, \citenamefont {Durand}, \citenamefont {Kemiktarak},\ and\ \citenamefont {Lawall}}]{stambaugh2014cavity}%
  \BibitemOpen
  \bibfield  {author} {\bibinfo {author} {\bibfnamefont {C.}~\bibnamefont {Stambaugh}}, \bibinfo {author} {\bibfnamefont {M.}~\bibnamefont {Durand}}, \bibinfo {author} {\bibfnamefont {U.}~\bibnamefont {Kemiktarak}}, \ and\ \bibinfo {author} {\bibfnamefont {J.}~\bibnamefont {Lawall}},\ }\bibfield  {title} {\enquote {\bibinfo {title} {Cavity-enhanced measurements for determining dielectric-membrane thickness and complex index of refraction},}\ }\href {https://opg.optica.org/ao/abstract.cfm?uri=ao-53-22-4930} {\bibfield  {journal} {\bibinfo  {journal} {Applied Optics}\ }\textbf {\bibinfo {volume} {53}},\ \bibinfo {pages} {4930--4938} (\bibinfo {year} {2014})}\BibitemShut {NoStop}%
\bibitem [{\citenamefont {Steinlechner}\ \emph {et~al.}(2017)\citenamefont {Steinlechner}, \citenamefont {Kr{\"u}ger}, \citenamefont {Martin}, \citenamefont {Bell}, \citenamefont {Hough}, \citenamefont {Kaufer}, \citenamefont {Rowan}, \citenamefont {Schnabel},\ and\ \citenamefont {Steinlechner}}]{steinlechner2017optical}%
  \BibitemOpen
  \bibfield  {author} {\bibinfo {author} {\bibfnamefont {J.}~\bibnamefont {Steinlechner}}, \bibinfo {author} {\bibfnamefont {C.}~\bibnamefont {Kr{\"u}ger}}, \bibinfo {author} {\bibfnamefont {I.~W.}\ \bibnamefont {Martin}}, \bibinfo {author} {\bibfnamefont {A.}~\bibnamefont {Bell}}, \bibinfo {author} {\bibfnamefont {J.}~\bibnamefont {Hough}}, \bibinfo {author} {\bibfnamefont {H.}~\bibnamefont {Kaufer}}, \bibinfo {author} {\bibfnamefont {S.}~\bibnamefont {Rowan}}, \bibinfo {author} {\bibfnamefont {R.}~\bibnamefont {Schnabel}}, \ and\ \bibinfo {author} {\bibfnamefont {S.}~\bibnamefont {Steinlechner}},\ }\bibfield  {title} {\enquote {\bibinfo {title} {Optical absorption of silicon nitride membranes at 1064 nm and at 1550 nm},}\ }\href {https://journals.aps.org/prd/abstract/10.1103/PhysRevD.96.022007} {\bibfield  {journal} {\bibinfo  {journal} {Physical Review D}\ }\textbf {\bibinfo {volume} {96}},\ \bibinfo {pages} {022007} (\bibinfo {year} {2017})}\BibitemShut {NoStop}%
\bibitem [{\citenamefont {Nejadriahi}\ \emph {et~al.}(2020)\citenamefont {Nejadriahi}, \citenamefont {Friedman}, \citenamefont {Sharma}, \citenamefont {Pappert}, \citenamefont {Fainman},\ and\ \citenamefont {Yu}}]{nejadriahi2020thermo}%
  \BibitemOpen
  \bibfield  {author} {\bibinfo {author} {\bibfnamefont {H.}~\bibnamefont {Nejadriahi}}, \bibinfo {author} {\bibfnamefont {A.}~\bibnamefont {Friedman}}, \bibinfo {author} {\bibfnamefont {R.}~\bibnamefont {Sharma}}, \bibinfo {author} {\bibfnamefont {S.}~\bibnamefont {Pappert}}, \bibinfo {author} {\bibfnamefont {Y.}~\bibnamefont {Fainman}}, \ and\ \bibinfo {author} {\bibfnamefont {P.}~\bibnamefont {Yu}},\ }\bibfield  {title} {\enquote {\bibinfo {title} {Thermo-optic properties of silicon-rich silicon nitride for on-chip applications},}\ }\href {https://opg.optica.org/oe/fulltext.cfm?uri=oe-28-17-24951&id=434318} {\bibfield  {journal} {\bibinfo  {journal} {Optics Express}\ }\textbf {\bibinfo {volume} {28}},\ \bibinfo {pages} {24951--24960} (\bibinfo {year} {2020})}\BibitemShut {NoStop}%
\bibitem [{\citenamefont {Sadeghi}\ \emph {et~al.}(2020)\citenamefont {Sadeghi}, \citenamefont {Demir}, \citenamefont {Villanueva}, \citenamefont {K{\"a}hler},\ and\ \citenamefont {Schmid}}]{sadeghi2020frequency}%
  \BibitemOpen
  \bibfield  {author} {\bibinfo {author} {\bibfnamefont {P.}~\bibnamefont {Sadeghi}}, \bibinfo {author} {\bibfnamefont {A.}~\bibnamefont {Demir}}, \bibinfo {author} {\bibfnamefont {L.~G.}\ \bibnamefont {Villanueva}}, \bibinfo {author} {\bibfnamefont {H.}~\bibnamefont {K{\"a}hler}}, \ and\ \bibinfo {author} {\bibfnamefont {S.}~\bibnamefont {Schmid}},\ }\bibfield  {title} {\enquote {\bibinfo {title} {Frequency fluctuations in nanomechanical silicon nitride string resonators},}\ }\href {https://journals.aps.org/prb/abstract/10.1103/PhysRevB.102.214106} {\bibfield  {journal} {\bibinfo  {journal} {Physical Review B}\ }\textbf {\bibinfo {volume} {102}},\ \bibinfo {pages} {214106} (\bibinfo {year} {2020})}\BibitemShut {NoStop}%
\bibitem [{\citenamefont {Zhang}\ and\ \citenamefont {St-Gelais}(2023)}]{zhang2023demonstration}%
  \BibitemOpen
  \bibfield  {author} {\bibinfo {author} {\bibfnamefont {C.}~\bibnamefont {Zhang}}\ and\ \bibinfo {author} {\bibfnamefont {R.}~\bibnamefont {St-Gelais}},\ }\bibfield  {title} {\enquote {\bibinfo {title} {Demonstration of frequency stability limited by thermal fluctuation noise in silicon nitride nanomechanical resonators},}\ }\href {https://pubs.aip.org/aip/apl/article/122/19/193501/2889989/Demonstration-of-frequency-stability-limited-by} {\bibfield  {journal} {\bibinfo  {journal} {Applied Physics Letters}\ }\textbf {\bibinfo {volume} {122}} (\bibinfo {year} {2023})}\BibitemShut {NoStop}%
\bibitem [{\citenamefont {Beccari}\ \emph {et~al.}(2022)\citenamefont {Beccari}, \citenamefont {Visani}, \citenamefont {Fedorov}, \citenamefont {Bereyhi}, \citenamefont {Boureau}, \citenamefont {Engelsen},\ and\ \citenamefont {Kippenberg}}]{beccari2022strained}%
  \BibitemOpen
  \bibfield  {author} {\bibinfo {author} {\bibfnamefont {A.}~\bibnamefont {Beccari}}, \bibinfo {author} {\bibfnamefont {D.~A.}\ \bibnamefont {Visani}}, \bibinfo {author} {\bibfnamefont {S.~A.}\ \bibnamefont {Fedorov}}, \bibinfo {author} {\bibfnamefont {M.~J.}\ \bibnamefont {Bereyhi}}, \bibinfo {author} {\bibfnamefont {V.}~\bibnamefont {Boureau}}, \bibinfo {author} {\bibfnamefont {N.~J.}\ \bibnamefont {Engelsen}}, \ and\ \bibinfo {author} {\bibfnamefont {T.~J.}\ \bibnamefont {Kippenberg}},\ }\bibfield  {title} {\enquote {\bibinfo {title} {Strained crystalline nanomechanical resonators with quality factors above 10 billion},}\ }\href {https://www.nature.com/articles/s41567-021-01498-4} {\bibfield  {journal} {\bibinfo  {journal} {Nature Physics}\ }\textbf {\bibinfo {volume} {18}},\ \bibinfo {pages} {436--441} (\bibinfo {year} {2022})}\BibitemShut {NoStop}%
\bibitem [{\citenamefont {Liu}\ \emph {et~al.}(2011)\citenamefont {Liu}, \citenamefont {Usami}, \citenamefont {Naesby}, \citenamefont {Bagci}, \citenamefont {Polzik}, \citenamefont {Lodahl},\ and\ \citenamefont {Stobbe}}]{liu2011high}%
  \BibitemOpen
  \bibfield  {author} {\bibinfo {author} {\bibfnamefont {J.}~\bibnamefont {Liu}}, \bibinfo {author} {\bibfnamefont {K.}~\bibnamefont {Usami}}, \bibinfo {author} {\bibfnamefont {A.}~\bibnamefont {Naesby}}, \bibinfo {author} {\bibfnamefont {T.}~\bibnamefont {Bagci}}, \bibinfo {author} {\bibfnamefont {E.~S.}\ \bibnamefont {Polzik}}, \bibinfo {author} {\bibfnamefont {P.}~\bibnamefont {Lodahl}}, \ and\ \bibinfo {author} {\bibfnamefont {S.}~\bibnamefont {Stobbe}},\ }\bibfield  {title} {\enquote {\bibinfo {title} {High-q optomechanical gaas nanomembranes},}\ }\href {https://pubs.aip.org/aip/apl/article-abstract/99/24/243102/122224/High-Q-optomechanical-GaAs-nanomembranes} {\bibfield  {journal} {\bibinfo  {journal} {Applied Physics Letters}\ }\textbf {\bibinfo {volume} {99}} (\bibinfo {year} {2011})}\BibitemShut {NoStop}%
\bibitem [{\citenamefont {Cole}\ \emph {et~al.}(2014)\citenamefont {Cole}, \citenamefont {Yu}, \citenamefont {G{\"a}rtner}, \citenamefont {Siquans}, \citenamefont {Moghadas~Nia}, \citenamefont {Schm{\"o}le}, \citenamefont {Hoelscher-Obermaier}, \citenamefont {Purdy}, \citenamefont {Wieczorek}, \citenamefont {Regal} \emph {et~al.}}]{cole2014tensile}%
  \BibitemOpen
  \bibfield  {author} {\bibinfo {author} {\bibfnamefont {G.~D.}\ \bibnamefont {Cole}}, \bibinfo {author} {\bibfnamefont {P.-L.}\ \bibnamefont {Yu}}, \bibinfo {author} {\bibfnamefont {C.}~\bibnamefont {G{\"a}rtner}}, \bibinfo {author} {\bibfnamefont {K.}~\bibnamefont {Siquans}}, \bibinfo {author} {\bibfnamefont {R.}~\bibnamefont {Moghadas~Nia}}, \bibinfo {author} {\bibfnamefont {J.}~\bibnamefont {Schm{\"o}le}}, \bibinfo {author} {\bibfnamefont {J.}~\bibnamefont {Hoelscher-Obermaier}}, \bibinfo {author} {\bibfnamefont {T.~P.}\ \bibnamefont {Purdy}}, \bibinfo {author} {\bibfnamefont {W.}~\bibnamefont {Wieczorek}}, \bibinfo {author} {\bibfnamefont {C.~A.}\ \bibnamefont {Regal}},  \emph {et~al.},\ }\bibfield  {title} {\enquote {\bibinfo {title} {Tensile-strained inxga1- xp membranes for cavity optomechanics},}\ }\href {https://pubs.aip.org/aip/apl/article-abstract/104/20/201908/130788} {\bibfield  {journal} {\bibinfo  {journal} {Applied Physics Letters}\ }\textbf {\bibinfo {volume} {104}} (\bibinfo {year}
  {2014})}\BibitemShut {NoStop}%
\bibitem [{\citenamefont {Xu}\ \emph {et~al.}(2023)\citenamefont {Xu}, \citenamefont {Shin}, \citenamefont {Sberna}, \citenamefont {van~der Kolk}, \citenamefont {Cupertino}, \citenamefont {Bessa},\ and\ \citenamefont {Norte}}]{xu2023high}%
  \BibitemOpen
  \bibfield  {author} {\bibinfo {author} {\bibfnamefont {M.}~\bibnamefont {Xu}}, \bibinfo {author} {\bibfnamefont {D.}~\bibnamefont {Shin}}, \bibinfo {author} {\bibfnamefont {P.~M.}\ \bibnamefont {Sberna}}, \bibinfo {author} {\bibfnamefont {R.}~\bibnamefont {van~der Kolk}}, \bibinfo {author} {\bibfnamefont {A.}~\bibnamefont {Cupertino}}, \bibinfo {author} {\bibfnamefont {M.~A.}\ \bibnamefont {Bessa}}, \ and\ \bibinfo {author} {\bibfnamefont {R.~A.}\ \bibnamefont {Norte}},\ }\bibfield  {title} {\enquote {\bibinfo {title} {High-strength amorphous silicon carbide for nanomechanics},}\ }\href {https://onlinelibrary.wiley.com/doi/abs/10.1002/adma.202306513} {\bibfield  {journal} {\bibinfo  {journal} {Advanced Materials}\ ,\ \bibinfo {pages} {2306513}} (\bibinfo {year} {2023})}\BibitemShut {NoStop}%
\end{thebibliography}%
\end{document}